 \documentclass[final,3p,times,authoryear]{elsarticle}

\usepackage{amssymb}
\usepackage{amsthm}

\usepackage[caption=false]{subfig}

\usepackage{ulem,lipsum}
\usepackage{soul}
\usepackage{xcolor}
\usepackage{mathtools}
\usepackage{amsmath}
\mathtoolsset{showonlyrefs}

\usepackage[hidelinks,colorlinks=false]{hyperref}

\newtheorem{lemma}{Lemma}

\newtheorem{remark}{Remark}
\newtheorem{definition}{Definition}
\newtheorem{existing result}{Existing Result}
\journal{Transportation Research: Part B}

\begin{document}

\begin{frontmatter}



\title{Modeling Information Propagation in General V2V-enabled Transportation Networks}

\author[add1]{Jungyeol Kim\corref{cor1}}
\ead{jungyeol@seas.upenn.edu}
\author[add1]{Saswati Sarkar}
\ead{swati@seas.upenn.edu}
\author[add1]{Santosh S. Venkatesh}
\ead{venkates@seas.upenn.edu}
\author[add1,add2]{Megan Smirti Ryerson}
\ead{mryerson@design.upenn.edu}
\author[add3]{David Starobinski}
\ead{staro@bu.edu}

\cortext[cor1]{Corresponding author}

\address[add1]{Department of Electrical and Systems Engineering, University of Pennsylvania, Philadelphia, PA 19104, United States}
\address[add2]{Department of City and Regional Planning, University of Pennsylvania, Philadelphia, PA 19104, United States}
\address[add3]{Division of Systems Engineering, Boston University, Boston, MA, 02215, Unites States}


\begin{abstract}
V2V technologies bridge two infrastructures: the communications infrastructure and the transportation infrastructure. These infrastructures are interconnected and interdependent. On the one hand, the communications network enables V2V interactions, while, on the other hand, the density of vehicles on the roadway enabled with V2V and the level of congestion on the roadway determine the speed and quality of communications between vehicles and infrastructure. The V2V technology is expected to contribute significantly to the growth of shared mobility, in turn, receives a significant boost from the deployment of a large number of connected vehicles in shared mobility services, provided challenges towards the deployment can be overcome. Vehicle mobility patterns and communication conditions are not only heterogeneous, but they also evolve constantly, leading to dynamic coupling between the communication and the transportation infrastructure. We consider the communication of messages amongst the vehicles in a transportation network, and estimate how quickly messages spread under different conditions of traffic density (traffic congestion, the presence of an accident, and time of day such as morning and evening rush hour) and communication conditions. We developed a continuous-time Markov chain to describe the information propagation process through enabled vehicles. Our models converge to a solution of a set of clustered epidemiological differential equations which lend itself to fast computation. We then demonstrate the applicability of this model in various scenarios: both real-world scenarios and hypothesized scenarios of outages and system perturbations. We find that our models match actual trajectory data with very little error, demonstrating the applicability of our models to study the spread of information through a network of connected vehicles.
\end{abstract}


\end{frontmatter}


\section{Introduction}\label{intro}

Emerging Vehicle-to-Vehicle (V2V) technology is poised to significantly impact the functioning and management of transportation networks~\citep{harding2014vehicle,v2v_comm2}. With V2V, vehicles can communicate directly with each other to pass on information about road conditions ahead. Vehicles may also communicate with bikes, wheelchairs, and devices held by pedestrians. We consider all these communications within the purview of V2V. The primary goal of V2V is to improve vehicle safety by providing warnings to drivers about dangerous situations. Applications of V2V include forward collision warning, blind spot warning, intersection movement assist, and left turn assist; the latter two applications alone could prevent an estimated 600,000 crashes and save over 1,000 lives each year in the U.S.~\cite{bertini2016assessing}. Automakers such as General Motors and Toyota are currently researching and incorporating V2V technology in vehicles today, with Toyota announcing plans to adopt the technology for all cars in their US market by 2021 \citep{toyota}. 

In order for vehicles to benefit from V2V communication, the number of connected vehicles must reach a critical mass. As the number of V2V-enabled vehicles on the road increases, the safety benefits provided by V2V technology will be enhanced. Improved benefits could create a positive feedback loop, leading to the deployment of more V2V-enabled vehicles. Widespread adoption of this technology, however, will not likely start with individual consumers. Individual consumers tend to hold on to personal vehicles for many years despite new technology developments; consider that the average age of cars and light trucks reached 11.6 years in 2016 \citep{age_vehicle}, an increase over previous years. The early adopters will most likely be shared passenger transportation services – such as Uber and Lyft -- as well as burgeoning shared freight companies, which will see large-scale safety and efficiency benefits from deploying a shared connected vehicle fleet. 

Shared freight services have significant potential for savings if individual fleets collaborate and communicate. Information sharing enables freight-sharing firms to use multiple truck convoys to optimize cargo deliveries, leading to savings; consider that \cite{fernandez2018shared} finds that when carriers collaborate and share customers, freight companies can save on costs of up to 25\%. V2V communication will allow for not just the sharing of information on where and when to pick up deliveries, but for information on traffic flow and safety hazards. With a wide-scale deployment of V2V-enabled trucks, shared freight vehicles could also save fuel; through V2V communication, vehicles can optimize when and how to converge on a network and platoon maintaining a constant distance between vehicles, thus optimizing fuel efficiency and ensuring safety (which can save up to 20\% of fuel consumption per~\cite{adler2016optimal,mitnews}). 

On the shared passenger mobility side, shared mobility companies are already connected through the wireless communications network. Drivers of shared vehicles are connected to their home company’s systems and to traffic information systems such as Waze, while individual passengers are connected via their mobile phones. Shared passenger mobility companies are also on the forefront of new technologies, with Uber investing heavily in autonomous vehicle (AV) research and development. V2V is critical in AV development and deployment; consider that the crash of an autonomous Uber vehicle in Tempe, Arizona in March 2018 showed that the vehicle did not show any signs of slowing down before the fatal accident \citep{uber1}. If V2V technology had been deployed in this vehicle, there may have been a warning about the imminent collision \citep{uber2}. AV sensors detect objects that are currently visible, while V2V offers the ability to detect potential dangers in limited sight situations and avoid risks in advance. 

Not only is V2V expected to facilitate the growth of shared mobility industry, but it also provides a mechanism for the connected vehicles in shared transportation services to enhance the societal objectives. Specifically, owing to the sheer number of vehicles in shared mobility services, they can exchange information regarding safety and traffic conditions not only among themselves but with other vehicles which do not belong to shared mobility services. In this paper, we explore the propagation of messages – which could comprise of utilitarian messages such as information about upcoming hazards or traffic conditions – through a fleet of connected vehicles. While we envision these vehicles to be a shared fleet in the near term, as individuals adopt V2V-enabled vehicles they will become part of the information sharing fleet in the long term. The models we present are flexible in the vehicles that comprise the fleet, rendering them applicable for the near term with shared fleets and the long term with mixed shared fleets and individual vehicles.

\subsection{Challenges}
A key challenge in V2V networks lies in characterizing the speed (or conversely the latency) of information propagation between vehicles, which are metrics of utmost importance for safety applications. Indeed, fundamental characteristics of vehicular networks complicate the task of obtaining a mathematically tractable model that is computationally simple and captures key attributes. Both vehicle mobility and wireless communication influence information propagation in the vehicular network, and both conditions vary temporally and spatially. In the former, we observe that the mobility of vehicles is heterogeneous because of the following factors: (1) various forms of a transportation network such as grid road network (e.g., Manhattan) \citep{marcuse1987grid}, radial and circular road topologies (e.g., Paris, Moscow) \citep{badia2014competitive} and irregular narrow streets; (2) a regional characteristic, such as urban and rural; (3) time of day, such as morning and evening rush hour. In the latter, the heterogeneity of communication conditions can primarily be attributed to the followings: (1) fading due to obstacles such as buildings and trees; (2) fading due to adverse weather conditions such as heavy storms; (3) frequency of communication. These various forms of transportation and communication conditions have different consequences for dynamics of information propagation. Because of the above, the model should be general in that it can capture arbitrary transportation and communication conditions. 

In addition to the heterogeneity of the transportation and communication networks, both mobility patterns and communication continuously evolve over space and time. Vehicle mobility is constantly changing because of (1) evolving road topologies (e.g., due to road constructions and temporary roadblocks) and (2) varying traffic conditions (e.g., due to congestion and accidents). Communication conditions may also change over time (e.g., due to fluctuating link quality). The evolution of mobility and communication conditions alter information propagation dynamics. In order to reflect the changes over time, any model should be readily generalizable depending on the evolution of mobility patterns and communication conditions.

Finally, any model developed must lend itself to simple computation regardless of scale. Consider that the number of vehicles registered in Los Angeles in 2017 was around 8 million~\citep{car_LA} and the number of registered vehicles in Shanghai in 2015 was about 2.5 million~\citep{car_shanghai}. 
Given this sheer number of vehicles and considering also non-motorized travelers, it is necessary that computation gracefully scales with the number of entities.

\subsection{Contributions}

Prior research on V2V has focused on how to leverage this technology to smooth out congestion speeds~\citep{scholliers2016impact} (transportation problem with communication exogenous) and on the relationship between the latency of different communication mechanisms and their impact on traffic flow and on vulnerable road users~\citep{whyte2013security,zhang2014defending,wired15} (communications problem with transportation exogenous). While message propagation speed in vehicular network has also been investigated, many studies focus only on movements along a one-dimensional road~\citep{agarwal2012phase,kesting2010connectivity,baccelli2012highway,zhang2014stochastic}. A recent study investigated the delay to forward messages in a two-dimensional road and introduced an algorithm for selecting the path with the lowest expected delay~\citep{he2017delay}. Others have studied the metapopulation spreading process in which a group of individuals is viewed as a node and individuals move along link connecting nodes~\citep{levins1969some,hanski2004ecology,sattenspiel1995structured,keeling2002estimating,gallos2004absence,watts2005multiscale,shausan2015model,sani2007stochastic}. 

However, studies that use actual microscopic trajectory data to validate the information propagation model are scant. Additionally, no prior study seems to account for arbitrary road topologies and temporal variations of traffic density and routing, thus motivating our work on modeling dynamics of V2V information propagation under various vehicle movement patterns and communication conditions on arbitrary road networks.
Following on numerous studies that use Markov chains to estimate freeway travel time in both routine and perturbed states~\citep{ramezani2012estimation,dong2009flow,geroliminis2005prediction,alfa1995modelling}, we use a continuous-time Markov chain (CTMC) to describe the information propagation process.
CTMC is known to be computationally challenging particularly when the number of vehicles and roads increase. A key insight of our work is that by applying the law of large numbers, the propagation process can be shown to converge to a solution of a set of clustered epidemiological differential equations which lend itself to fast computation. To the best of our knowledge, this is the first work that uses clustered epidemiological differential equations in vehicular networks. 

In this work, we first show how these differential equations can be used to model a wide variety of transportation and communication conditions, based only on an appropriate choice of parameters. Next, we show that the model is easily generalizable even when transportation and communication conditions change over time. Moreover, the computation time needed to solve the equations does not depend on the number of entities including vehicles, pedestrians, bikes, and wheelchairs.

Next, we conduct detailed empirical verification of the model. This empirical verification specifically considers scenarios which are not within the purview of the mathematical results owing to analytical assumptions. Indeed, the theoretical model assumes that the propagation process converges to the solutions of differential equations (1) as the number of vehicles goes to infinity; and (2) as the vehicles follow exponential sojourn times at different locations.

The empirical evaluations, on the other hand, only consider a finite number of vehicles. In addition, we consider various statistical models, which we call synthetic models, with realistic mobility patterns and communication conditions. We also use extensive real-life vehicle trajectory data collected from various road topologies (highway, roads with intersections, roundabout). The empirical validation confirms that, even for a moderate number of vehicles, the output of the differential equation model closely matches the results of the propagation process. We stress that the real-life trajectory data do not satisfy the exponential sojourn time assumption and apply to a finite number of vehicles. Nonetheless, we show that our model leads to a solution that matches well simulation results based on actual trace data collected under various road topologies.

We subsequently show how our model can be utilized to answer various technical questions that may arise in a context of V2V information propagation. Specifically, we first show how traffic congestion can be alleviated through an intelligent application of V2V technology. We then assess how quickly information about the location of disruptive changes (i.e., temporary roadblocks) can be disseminated throughout a city via V2V technology. Recall that given a large number of vehicles participating in shared transportation such sharing economy can be instrumental in facilitating societal objectives of alleviating traffic congestion and disruption through the dissemination of relevant messages. Lastly, we examine how the initial location of informed vehicles determines the spread of information throughout the transportation network. Our examination reveals a counter-intuitive phenomenon, that is, message propagation is not necessarily accelerated if the initially informed vehicles are centrally located.

The rest of the paper consists of the following. In Section \ref{sec:2}, we introduce our model based on clustered epidemiological differential equations. The section also includes a mathematical justification of the results and describes generalized statistical models applicable to various mobility patterns and communication conditions. In Section \ref{sec:3}, we validate the applicability of our modeling methodology using synthetic models and real-life traffic trace. In Section \ref{sec:4}, we present case-studies demonstrating the usefulness of our model to characterize the behavior of V2V systems in various practical situations. In Section \ref{sec:5}, we show that our model is computationally tractable regardless of the number of vehicles. Section \ref{sec:6} summarizes our research findings.

\section{Model Formulation} \label{sec:2}

Section 2 introduces models that reflect the various forms of vehicle mobility and communication that exist in the real world. We first introduce a general mathematical framework for the propagation of messages in V2V systems (Section \ref{ssec:2.A}). We subsequently show how the general framework caters to various specialized cases that arise in practice (Section \ref{ssec:2.2}): temporal variation of the traffic density and routing to capture the vehicle movement pattern during rush hour (Section \ref{sssec:2.B.1}), a location dependent mobility model that reflects speed limits applied differently depending on the region (Section \ref{sssec:2.B.2}), and finally a traffic density dependent mobility model that reflects reduced vehicle speed due to high traffic density (Section \ref{sssec:2.B.3}). These various scenarios can occur simultaneously, which can easily be represented by combining these models. All models are based on continuous time Markov chains. We show that information propagation based on the Markov chain can be very well approximated by differential equations for the various realistic vehicle mobility and communication patterns mentioned above.

\subsection{Clustered Epidemiological Differential Equation based model} \label{ssec:2.A}

\begin{figure} 
	\centering
	\subfloat[Grid roads]{%
		\includegraphics[width=.2\linewidth]{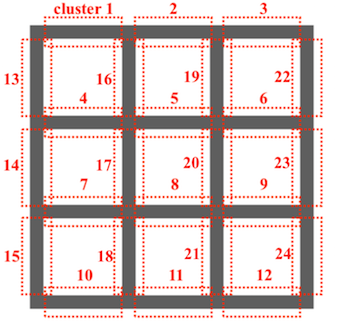}%
	}%
	\hspace{5pt}
	\subfloat[Radial and circular roads]{%
		\includegraphics[width=.2\linewidth]{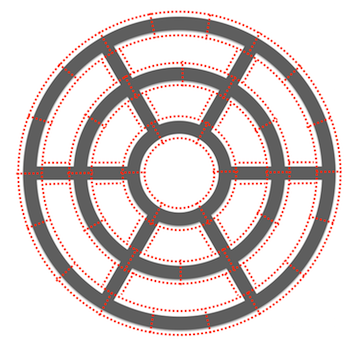}%
	}%
	\hspace{5pt}
	\subfloat[Diamond interchange]{%
		\includegraphics[width=.21\linewidth]{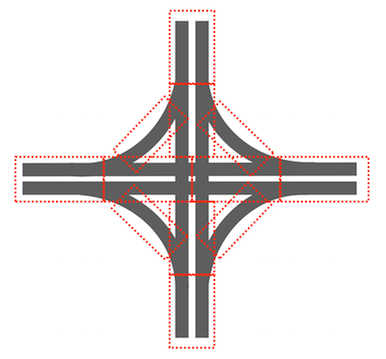}%
	}%
	\hspace{1pt}
	\subfloat[Irregular roads]{%
		\includegraphics[width=.24\linewidth]{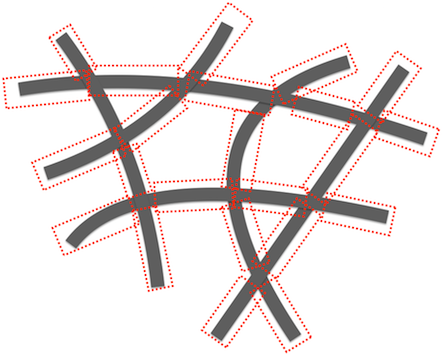}%
	}%
	\captionsetup{singlelinecheck = false, justification=raggedright, labelsep=space}
	\caption{Clustered road topologies. The figures represent various road topologies such as grid roads, radial with circular roads, and irregular roads. In arbitrary types of topologies, roads can be divided into multiple smaller segments, so clusters can be defined as shown by the red dotted rectangle in the illustration.}
	\label{fig:topologies}
\end{figure}

We develop tools to model information propagation under general types of transportation network and various communication conditions. Transportation networks exist in various forms such as highways connecting cities; coastal roads; and urban roads. Unlike highways or coastal roads, which are relatively simple one-dimensional forms, urban roads exist in complex networks of different types depending on the characteristics of the area. In most areas, however, these road topologies are superposed in a complex manner. We introduce mathematical tools that can be used to model and analyze information flow across arbitrary complex road networks as shown in Figure~\ref{fig:topologies}, which can potentially be used for prediction of information propagation through vehicle-to-vehicle communication.

The extent to which information is spread between moving vehicles is determined through vehicle movement and wireless communication. The mobility of vehicles on the transportation network depends on topology which will continuously evolve (addition of new roads, road blockage of existing roads due to maintenance), traffic conditions (traffic congestion, the presence of an accident), time of day, and the characteristics of the individual travelers (urban, rural, land use interactions). Communication on these transportation networks is influenced by traffic conditions (packet collisions due to high traffic density) and communication conditions (frequency of communication between vehicles, fading due to obstacles such as buildings, trees, etc.).
\begin{figure} 
	\centering
	\includegraphics[width=.9\linewidth]{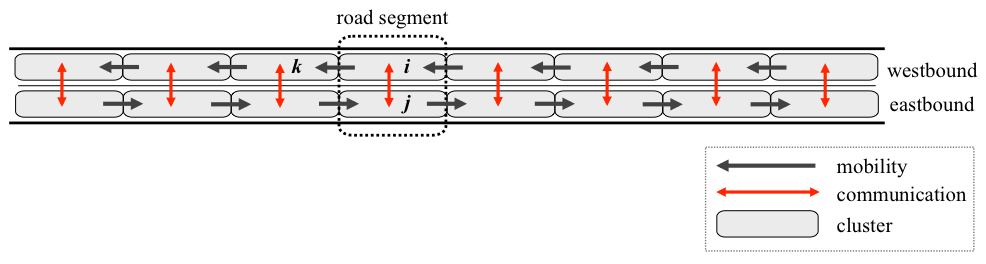}%
	\captionsetup{singlelinecheck = false, justification=raggedright, labelsep=space}
	\caption{Mobility and communication networks. The mobility network is a directed network and the communication network is an undirected network. The edges of these two networks may overlap but need not be exactly the same.}
	\label{fig:network}
\end{figure} 
We present a mathematical model that can capture information flow in arbitrary vehicular network that contains all of these complex elements.

We divide the entire road topology into a collection of $J$ clusters, with each cluster corresponding to a specific region of the road; thus each vehicle is located in one of the $J$ clusters. One possible way to form a cluster is to set the cluster size to the communication range as it is natural to assume that the vehicles located within the same cluster are within the V2V communication range. However, vehicles located at the boundary of one cluster can communicate with vehicles in other adjacent clusters if they are within the communication range. Our model can cater to this communication between vehicles located near the boundary of two adjacent clusters by considering that the vehicles located in the different clusters can communicate at a reduced rate. When considering the communication of vehicles in adjacent clusters, it is reasonable to apply the reduced rate because not all cars in the two adjacent clusters can communicate with each other and only vehicles located near the boundary can communicate. 

Vehicles can both communicate and move across clusters or within clusters. Clearly, a vehicle cannot communicate between every pair of clusters because of the technical limitations of the wireless communication range. Similarly, a vehicle cannot move between each pair of clusters due to the nature of the road and traffic rules. Both of the above are influenced, but not solely determined, by geography. For example, although two clusters are very close enough to permit communication, the vehicle may not be able to move between the two clusters. Figure~\ref{fig:network} shows cluster $i$ and $j$ located on the same road segment of the two-way roads; these vehicles are close enough to permit communication but traffic rules do not allow vehicles to travel across the median. Similarly, even between two adjacent clusters where vehicles can move, the success rate of communication between two clusters may become low or 0 due to obstacles (buildings, trees, etc.).

\begin{table}[htp]
	\centering	
	\caption{Mathematical Notation}
	\label{tab:notation}
	\begin{tabular}{l   l}
		\hline
		\\[-1.8ex] 
		$G(V,E)$ &\quad \quad Directed network of mobility on road topology \\ 
		$G'(V,E')$ &\quad \quad Undirected network of communication on road topology \\ 
		$V$ &\quad \quad Set of clusters, $|V|=J$ \\ 
		$E$ &\quad \quad Set of directed mobility edges \\ 
		$E'$ &\quad \quad Set of undirected communication edges \\ 
		$N$ &\quad \quad Total number of vehicles \\ 
		$n^I_j(t)$ &\quad \quad Number of informed vehicles in cluster $j$ at time $t$\\ 
		$n^S_j(t)$ &\quad \quad Number of non-informed vehicles in cluster $j$ at time $t$\\ 
		$X(t)$ &\quad \quad Continuous-time Markov process\\ 
		$X_N(t)$ &\quad \quad Scaled Markov process, $X(t)/N$\\ 
		$\lambda^I_{ij}(\cdot)$ &\quad \quad Mobility rate from cluster $i$ to $j$ for informed vehicles\\
		$\lambda^S_{ij}(\cdot)$ &\quad \quad Mobility rate from cluster $i$ to $j$ for non-informed vehicles\\
		$\lambda$ &\quad \quad Upper bound for mobility rate\\
		$\beta_{ij}/N$ &\quad \quad Communication rate between a vehicle located in cluster $i$ and a vechicle in $j$\\
		$N_G(j)$ &\quad \quad Neighborhood of cluster $j$; set of clusters connected from cluster $j$\\
		$p_{jk}$ &\quad \quad Probability that a vehicle in cluster $j$ move to $k$\\ 
		$\rho(t)$ &\quad \quad Proportion of informed vehicles at time $t$\\ 
		[+1.ex]
		\hline
	\end{tabular}
\end{table}

We define two networks and corresponding adjacency metrics: communication network and mobility network. While both depend on the geographical characteristics of the roads, they do not necessarily have to be the same. We first describe the mobility network. Let $G = (V,E)$ be the directed mobility network on the road topology, and the directed networks $G$ consist of set of nodes $V =\left\{1,2,...,J\right\}$ corresponding to clusters and set of mobility edge set $E$. If clusters $j,k\in V$ are adjacent roads and vehicle movement is possible from cluster $j$ to $k$, the directed edge is specified as the edge $e\in E$ from $j$ to $k$ (equivalently, e:=$j\rightarrow k$). The corresponding adjacency matrix of $G$ is the $J$ $\times$ $J$ matrix $A = (a_{jk})$ where $a_{jk} = 1$ if $j\rightarrow k \in E$ and $a_{jk} = 0$ otherwise. We now introduce the communication network. Let ${G'} = (V,{E'})$ be the undirected network of communication with the set of same nodes $V$ and set of wireless communication edge set ${E'}$. If two clusters $j,k\in V$ can directly communicate amongst each other then there exist an edge $e\in {E'}$ between $j$ and $k$ (equivalently, e:=$j\leftrightarrow k$). The corresponding adjacency matrix ${G'}$ is the $J$ $\times$ $J$ symmetric matrix ${a'} = ({a'}_{jk})$ where ${a'}_{jk} = {a'}_{kj} = 1$ if $j\leftrightarrow k \in {E'}$ and ${a'}_{jk} = {a'}_{kj} = 0$ otherwise. Through the adjacency matrices $G$ and $G'$ we have discussed, the characteristics of any road topology in arbitrary city can be extracted.

We model the information propagation in transportation networks based on the Susceptible-Infective(SI) epidemiological model that has been taken in numerous infectious disease and information propagation research. This epidemiological model assumes that susceptible individuals have not yet incurred the disease but are vulnerable to it, and susceptible individuals can become infected after receiving the disease through contact with infected individuals. These newly infected individuals can spread the disease to susceptible individuals. In this study, we use the mathematical formulation where the notion of infective entities is taken on by vehicles which carry information, and the notion of susceptible entities from the epidemiological model is taken on by vehicles which do not yet have the information. From hence, we call the vehicle that carries the information as informed vehicles, and vehicles that have not received the information as non-informed vehicles. 

Suppose that $N$ vehicles are located in the network with each vehicle in one of the $J$ clusters corresponding to different road segments. We will suppose for now that the network is closed and that there are no exogenous arrivals into, or departures from, the system. Let ${{n_j}}^I(t)$ and ${{n_j}}^S(t)$ represent the number of informed and non-informed vehicles in cluster $j \in V$ at time $t$. The $2J$-dimensional lattice vector
\begin{equation*}
  \bigl({\bf{n}}^I(t),{\bf{n}}^S(t)\bigr) 
    = \bigl(n^I_1(t), n^I_2(t), \dots, n^I_{J}(t);\: n^S_1(t), n^S_2(t), \dots, 
      n^S_{J}(t)\bigr)
\end{equation*}
then represents the instantaneous state of the system, semicolon and extra spacing added merely for visual separation of informed and non-informed vehicular counts in the various clusters. The state space on which we model the dynamics of information propagation accordingly is the set of lattice points in $\mathbb{Z}^J\times\mathbb{Z}^J$ satisfying
\begin{equation}
  S^N := \left\{ ({\bf{n}}^I,{\bf{n}}^S)\mid n^I_j\geq0,\;n^S_j\geq 0,\; j=1,\dots,J;\;
    \sum_{j=1}^{J} \left(n^I_j + n^S_j\right) = N  \right\}.
\end{equation}
The basic transitions in this state space capture one of three types of phenomena: the movement of an informed vehicle to a neighbouring cluster; the movement of a non-informed vehicle to a neighbouring cluster; and the conversion of a non-informed vehicle to an informed vehicle by the successful transmission and receipt of information. A little notation helps grease the wheels: write ${\bf k} = ({\bf n}^I, {\bf n}^S)$ for the current state and let ${\bf 1}_j$ represent the $2J$-dimensional unit vector whose $j$th element is $1$ with all other elements being $0$. For $j$, $k\in\{1,\dots,J\}$ with $j\neq k$, the state transition ${\bf k} \rightarrow {\bf k} - {\bf 1}_j + {\bf 1}_k$ captures the movement of an informed vehicle from cluster $j$ to cluster $k$; the state transition ${\bf k} \rightarrow {\bf k} - {\bf 1}_{J+j} + {\bf 1}_{J+k}$ represents the movement of a non-informed vehicle from cluster $j$ to cluster $k$; and the state transition ${\bf k}\rightarrow {\bf k} + {\bf 1}_{k} - {\bf 1}_{J+k}$ represents a successful communication of information to an uninformed vehicle in cluster $k$ which now joins the informed ranks in that cluster with a concomitant reduction of the non-informed ranks in that cluster.

We now develop the stochastic underpinnings of the time evolution of the state process $X(t) = \bigl({\bf{n}}^I(t),{\bf{n}}^S(t)\bigr)$. We model mobility delays by assuming that the time taken by a vehicle to move to a neighboring cluster is exponentially distributed with possibly state-dependent parameters. Likewise, we model communication delays, within and across clusters, by supposing that the time taken for a successful transmission of information from an informed vehicle to a non-informed vehicle is exponentially distributed, again with possibly state-dependent parameters. Under these assumptions, the state evolution process $X(t)$ forms a continuous time Markov chain (CTMC). We flesh out the structure of the CTMC in what follows.

We recall that the CTMC exhibits the following three types of state transitions: (1) an informed vehicle moves from cluster $j$ to cluster $k$, $k\neq j$; (2) a non-informed vehicle moves from cluster $j$ to cluster $k$, $k\neq j$; and (3) a non-informed vehicle in a cluster $k$ receives a successful transmission from an informed vehicle located in the same cluster or in a different cluster.

The first two types of transition capture vehicle mobility. Write $\lambda_{jk}^I(\cdot)$ for the rate at which informed vehicles from cluster $j$ migrate to cluster $k$, and $\lambda_{jk}^S(\cdot)$ for the rate at which non-informed vehicles migrate from cluster $j$ to cluster $k$. These rates may be the same but there is no cost in the model to assuming potentially different mobility rates for informed and non-informed vehicles and we may as well do so. We assume that both $\lambda_{jk}^I (\cdot)$ and $\lambda_{jk}^S (\cdot)$ are bounded functions of $\frac{1}{N}({\bf{n}}^I,{\bf{n}}^S)$ if $a_{jk}=1$ and are $0$ otherwise. In other words, the model permits mobility-based transitions only between neighboring clusters, the transition rates between neighboring clusters are permitted to vary boundedly across clusters as a function of both the (geographic location of) the clusters as well as the density of vehicles in the clusters, and these rates may depend on whether the vehicle is informed or non-informed.

The third type of state transition that we encounter deals with a successful communication of information from an informed vehicle to a non-informed vehicle resulting in the non-informed vehicle attaining informed status. We posit fixed, non-negative constants $\beta_{jj}$ and $\beta_{jk}$, for each $j$ and $k$, such that intra-cluster communications between vehicles in a cluster $j$ occur at rate $\beta_{jj}/N$ while inter-cluster transmissions of information from cluster $j$ to a distinct cluster $k$ occur at rate $\beta_{jk}/N$. (To keep from unnecessarily burdening notation, we suppose that $\beta_{jk} = 0$ if $a'_{jk} = 0$, that is to say, there is no direct communication across clusters not connected by a wireless communication link.) The model explicitly captures the phenomenon that communication rates diminish due to reductions in shared bandwidth as the number of vehicles in the clusters increase. There is an implicit ergodic model assumption here: effectively, we assume that the number of vehicles in each cluster is proportional to the total number $N$ of vehicles in the network where the proportion of the population that is captured within each cluster may be cluster-dependent---these cluster-dependent constants of proportionality may be folded into the specification of the parameters $\beta_{jj}$ and $\beta_{jk}$. This type of phenomenon is familiar in ergodic chains where, with a large population $N$, the occupancy in each cluster will be close to its expected value. The implicit mobility network modelling assumption here, of course, is that the model clusters represent settings in which road segments may be reasonably considered to have an ergodic character where, with a sufficiently large population of vehicles, each cluster sees a non-trivial vehicular occupancy. 

To summarize, state transitions in the Markov chain are governed by exponential processes, the transitions from a given state ${\bf k} = \bigl({\bf n}^I, {\bf n}^S\bigr)$ to a state ${\bf k}' = {\bf k} + {\bf h}$ occurring at a rate
\begin{align} \label{eqn:transition_rate} 
q\left( {\bf k}, {\bf k} + {\bf{h}} \right) =
\begin{cases}
  \lambda^I_{jk}\left( \frac{\bf k}{N}\right) \cdot n^I_j
    & \text{if ${\bf{h}}=- {\bf 1}_j + {\bf 1}_k$ and $j\neq k$,}\\[4pt]
  \lambda^S_{jk} \left(\frac{\bf k}{N}\right) \cdot n^S_j 
    & \text{if ${\bf{h}}=- {\bf 1}_{J+j} + {\bf 1}_{J+k}$ and $j\neq k$,}\\[4pt]
  \frac{\beta_{jk}}{N} \cdot  n^I_j \cdot n^S_k 
    & \text{if ${\bf{h}}= {\bf 1}_{k} - {\bf 1}_{J+k}$,}\\[4pt]
  0 & \text{otherwise.}
\end{cases}
\end{align}
This follows a well-worn pathway in the theory of continuous-time Markov chains. The key to a dramatic asymptotic simplification in our setting is that the transition rates given by~\eqref{eqn:transition_rate} have a certain density-dependent property which reduces considerations via the ergodic theorem to a system of ordinary differential equations in the continuum. 

Proceed to the continuum limit and introduce the set $E := \bigl\{({\bf{I}},{\bf{S}})\mid I_i\geq 0, S_i\geq 0, i=1,2,\dots,J;\; \sum_{i=1}^{J} (I_i+S_i) = 1 \bigr\}$ where, in the natural vector notation, we write $\bigl({\bf{I}},{\bf{S}}\bigr) = (I_1,I_2,\dots,I_J;\;S_1,S_2,\dots,S_J)$. The continuous analog of~\eqref{eqn:transition_rate} is a continuous function $f\left( {\bf x},{\bf{h}} \right)$ on $E\times \mathbb{Z}^{2J}$ given, for each ${\bf x} = ({\bf I}, {\bf S})\in E$ and ${\bf h}\in\mathbb{Z}^{2J}$, by
\begin{equation} \label{eqn:f_function1}
f\left({\bf x},{\bf{h}}\right) =
\begin{cases}
  \lambda^I_{jk}\left({\bf{x}}\right) \cdot I_j
    & \text{if ${\bf{h}}=- {\bf 1}_j + {\bf 1}_k$ and $j\neq k$,}\\[4pt]
  \lambda^S_{jk}\left({\bf{x}}\right) \cdot S_j
    & \text{if ${\bf{h}}=- {\bf 1}_{J+j} + {\bf 1}_{J+k}$ and $j\neq k$,}\\[4pt]
  \beta_{jk} \cdot I_j \cdot S_k
    & \text{if ${\bf{h}}= {\bf 1}_{k} - {\bf 1}_{J+k}$,}\\[4pt]
  0 & \text{otherwise.}
\end{cases}       
\end{equation}
The discrete formulation~\eqref{eqn:transition_rate} can be imbedded in the continuous formulation~\eqref{eqn:f_function1} by the simple observation that $q\left( {\bf k}, {\bf k} + {\bf{h}} \right) = N f\left(\frac{\bf k}{N},{\bf{h}} \right)$, hence the connection to a continuum density---in the nomenclature introduced by Kurtz, we say that the Markov-chain is \textit{density-dependent}. In such cases a very general theorem of Kurtz (Kurtz, 1970) asserts that, in the asymptotic limit as $N\to\infty$, state evolution in the CTMC may be represented by a system of ordinary differential equations.

Introduce the formal notation
\begin{equation} \label{eqn:lm_1}
\lim_{N\rightarrow \infty} \frac{{\bf{n}}^I (t)}{N}={{\bf{I}}(t)},\quad \lim_{N\rightarrow \infty} \frac{{\bf{n}}^S (t)}{N}={{\bf{S}}(t)}.
\end{equation}
The formal quantities ${\bf I}(t)$ and ${\bf S}(t)$ represent the asymptotic fraction of informed and non-informed vehicles, respectively, in each cluster. The following is the key consequence of Kurtz's theorem adapted to our model assumptions. 

{\it Under the conditions of our model, for a given choice of initial conditions $\bigl({\bf I}(0), {\bf S}(0)\bigr)$, the time-evolution, $\bigl({\bf I}(t), {\bf S}(t)\bigr)$, of the distribution of the asymptotic fraction of informed and non-informed vehicles across clusters is governed by the following system of ordinary differential equations:}
\begin{equation} \label{eqn:diff_eqn}
\begin{split} 
\dot{I}_j(t)&=-\sum_{k\neq j}^{J} \lambda^I_{jk}\left({\bf{I,S}}\right) \cdot I_j +  \sum_{k=1}^{J}\beta_{kj} \cdot I_k\cdot S_j    + \sum_{k\neq j}^{J} \lambda^I_{kj}\left({\bf{I,S}}\right)\cdot  I_k \qquad (j=1,2,\dots,J),\\
\dot{S}_j(t)&=-\sum_{k\neq j}^{J}\lambda^S_{jk}\left({\bf{I,S}}\right) \cdot S_j -   \sum_{k=1}^{J}\beta_{kj} \cdot I_k \cdot S_j    + \sum_{k\neq j}^{J}\lambda^S_{kj}\left({\bf{I,S}}\right) \cdot S_k \qquad (j=1,2,\dots,J).
\end{split}       
\end{equation}
This reduction to a system of differential equations is the jumping off point for our model analysis and a reader who is primarily interested in seeing applications of the model in diverse settings can begin with~\eqref{eqn:diff_eqn} and read on. The theoretically inclined reader who would like to see details of how Kurtz's theorem, adapted to our setting, results in~\eqref{eqn:diff_eqn} will find the analysis and proofs in the Appendix.

\subsection{Specialization} \label{ssec:2.2}
The general framework that is a set of differential equations~\eqref{eqn:diff_eqn} cater to several special cases that arise in practice. For this, we consider the grid road topology in Figure~\ref{fig:grid_distance} with six avenues and streets. In the center of the network is a representation of a Central Business District (CBD): the CBD acts as an attractor of trips from the surrounding locations in the city (periphery). In this network, we assume that all roads are two-way and allow vehicles to move in both directions, and a road segment consists of two clusters corresponding to the opposite directional roads.

\subsubsection{Temporal variation of the traffic density and routing} \label{sssec:2.B.1}
\begin{figure}[hpt]
	\centering
	\includegraphics[width=.45\linewidth]{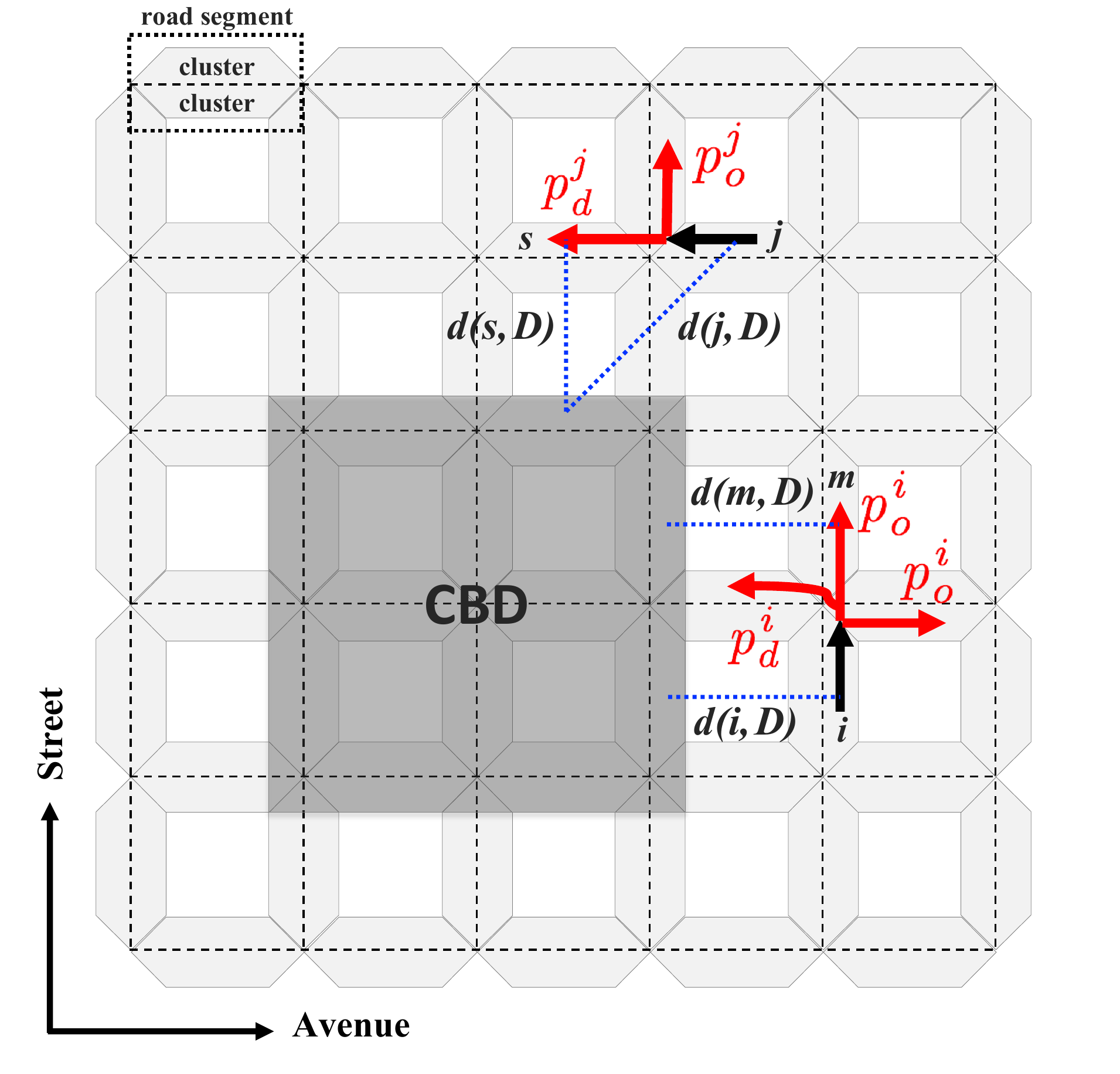}
	\captionsetup{singlelinecheck = false, justification=raggedright, labelsep=space}
	\caption{Clustered grid road topology. We capture probabilities of a vehicle moving between the CBD and the periphery. As for the movement from $j$ to $s$, the direction is classified as to the direction towards the CBD since $d(s,D)<d(j,D)$. Movement from $i$ to $m$ is classified as movement towards the periphery because the vehicle stays in the periphery, and does not move in the direction toward the CBD.} 
	\label{fig:grid_distance}
\end{figure} 
We now show how the differential equations~\eqref{eqn:diff_eqn} cater to the temporal variation of the traffic density and routing pertaining to vehicle movement during rush hour. We capture the morning rush from the periphery to the CBD in the morning and then the reverse at the conclusion of the work day. In the directed mobile network $G(V,E)$, the neighborhood of cluster $j$ is defined as the set of clusters connected from $j\in V$ through a directed edge, denoted $N_G(j)$. Let $D$ be the set of clusters in the CBD, and $O=D^c$ be the set of clusters in the periphery. As shown in Figure~\ref{fig:grid_distance}, the shaded area of the city center is generally located in the center of the city. Let $p_{jk}$ be the probability that vehicles in cluster $j$ move to cluster $k\in N_G(j)$. In that case, the mobility rate set in this model is
\begin{align}
\begin{split}
\lambda_{jk}^I(\cdot)=\lambda_{jk}^S(\cdot)={p_{jk}} \lambda
\label{eqn:temporal_rate}
\end{split}
\end{align}
where $\lambda$ is constant. Since the mobility rates for $j=1,2,...,J$ and $k\in N_G(j)$ are constant, it is clearly Lipschitz continuous on $E$. Therefore, by results in the Appendix, the behavior of propagation process based on this temporal variation of traffic density and routing can be approximated by ordinary differential equations \eqref{eqn:diff_eqn}. 

In the rest of this subsection, we describe how $p_{jk}$ can be computed. Suppose that, for the starting point $j\in \{1,2,...,J\}$, this probability is classified into two types: the probability $p^j_{d}$ of moving towards the CBD, and the probability $p^j_{o}$ of moving towards the periphery. Concretely, if the direction from cluster $j$ to $k$ is in the direction moving toward the periphery, $p_{jk}$ corresponds to $p^j_{o}$, and in the opposite case, $p_{jk}$ corresponds to $p^j_{d}$. We introduce the $\gamma$ parameter defined by the ratio of probability $p^j_{d}$ to $p^j_{o}$ to control the temporal variation of the routing for these two directions, resulting in $p^j_{d}=\gamma p^j_{o}$. For example, if $\gamma = 5$, the probability of moving towards the CBD is five times higher than the probability of moving towards the periphery. Thus, $\gamma>1$ indicates a movement pattern in which peripheral dwellers move into the city center, e.g., during morning commute time; $0<\gamma<1$ indicates a movement pattern in which workers leave the CBD and go back to the periphery, e.g., during evening commute time.

We define a decision rule for determining whether movement from one cluster to another cluster is toward or away from the CBD. Let $d(x, y)$ denote the Euclidean distance between two geometric centers of $x\in V$ and $y\in V$, and let $d(x, D)$ denote the minimum distance between a cluster $x\in V$ and a cluster in $D$ as follows.
\begin{align}
d(x,D) = \min\{ d(x,y)~|~y\in D\}
\end{align}
When vehicles move from cluster $j$ to $k$, if $d(j, D) < d(k,D)$ then it is classified as the movement towards the periphery, resulting in probability $p_{jk}$ being $p^j_{o}$. If $d(j, D) > d(k,D)$, it is classified as the movement towards the CBD, and as a result, the probability $p_{jk}$ becomes $p^j_{d}=\gamma p^j_{o}$. In the case of $d(j, D) = d(k,D)$, the type of movement direction is classified according to the position of the origin cluster and the destination cluster. Specifically, if $d(j, D) = d(k,D)$ and origin cluster $j$ and destination cluster $k$ are both in the CBD, $p_{jk}$ is considered to be $p^j_{d}=\gamma p^j_{o}$, since this direction corresponds to the movement to stay in the CBD rather than leaving out the CBD. Similarly, if $d(j, D) = d(k,D)$ and origin cluster $j$ and destination cluster $k$ are both in the periphery, $p_{jk}$ is regarded as $p^j_{o}$ because it is a type of movement that stays in the periphery, not in the direction toward the CBD.
Therefore, for all cluster $j\in V$ and its neighborhood $k\in N_G(j)$, we have
\begin{equation}
p_{jk} =
\begin{cases}
p^j_{d},\quad {\text {if}}~d(j, D) > d(k,D)\\
p^j_{d},\quad {\text {if}}~d(j, D) = d(k,D) {\text{ and }} j,k\in D\\
p^j_{o},\quad {\text {if}}~d(j, D) < d(k,D)\\
p^j_{o},\quad {\text {if}}~d(j, D) = d(k,D) {\text{ and }} j,k\in O
\label{decision-rule}
\end{cases}       
\end{equation}
From this decision rule, for cluster $j\in V$ and $k\in N_G(j)$, $p_{jk}$ is classified into one of $p^j_{d}$ and $p^j_{o}$, then these two probabilities $p^j_{d}$ and $p^j_{o}$ are determined by $\sum_{k\in N_G(j)} p_{jk}=1$. More specifically, we now show how $p^j_{d}$ and $p^j_{o}$ can be computed for all $j\in V$. First, consider an arbitrary node $j\in D$ and let $A$ be the set of clusters $k\in N_G(j)$ such that $d(j,D)\geq d(k,D)$. By using $\sum_{k\in N_G(j)} p_{jk}=1$, we have $|A|p_d^j+ |N_G(j)\backslash A|p_o^j=|A|\gamma p_o^j+ |N_G(j)\backslash A|p_o^j=1$, resulting in $p_o^j=\frac{1}{|A|\gamma+|N_G(j)\backslash A|}$ and $p_d^j=\frac{\gamma}{|A|\gamma+|N_G(j)\backslash A|}$. Similarly, for arbitrary node $j\in O$, let $A'$ be the set of clusters $k\in N_G(j)$ such that $d(j,D)\leq d(k,D)$. Through the same approach, we have $p_o^j=\frac{1}{|A'|+|N_G(j)\backslash A'|\gamma}$ and $p_d^j=\frac{\gamma}{|A'|+|N_G(j)\backslash A'|\gamma}$. For example, in Figure~\ref{fig:grid_distance}, when a vehicle in cluster $i$ moves to neighborhood clusters $k$, $n$, and $m$, the $p_{ik}$, $p_{in}$, and $p_{im}$ are classified as $p^i_o$, $p^i_d$, and $p^i_o$, respectively. Therefore, $p^i_o=1/(\gamma + 2)$ and $p^i_d=\gamma/(\gamma + 2)$, resulting in $p_{ik}=1/(\gamma + 2)$, $p_{in}=\gamma/(\gamma + 2)$, and $p_{im}=1/(\gamma + 2)$.
\subsubsection{Location dependent mobility model} \label{sssec:2.B.2}
In this subsection, we show that how the differential equations~\eqref{eqn:diff_eqn} can capture location dependent mobility model that reflects different speed limits for various regions. Speed limits are applied differently depending on the local characteristics of each city and the type of road, and as a result the average speed of the vehicles will depend on these characteristics. In this model, two different mobility rates are applied to reflect different speed limits applied to the CBD and the periphery roads, respectively, and the mobility rate of the CBD is set to a lower value. Let $\lambda_d$ be the mobility rate from a cluster located in the CBD to the neighbor clusters, so that the average time to stay in the cluster before moving to the neighbor cluster is 1 / $\lambda_d$. Similarly, let $\lambda_o$ be the mobility rate from a cluster located in the periphery to the neighbor clusters, so that the average time to stay in the cluster before moving to the neighbor cluster is 1 / $\lambda_o$. The routing probability associated with $\gamma$ is also applied. The location dependent mobility rate can be described as
\begin{align}
\begin{split}
\lambda_{jk}^I(\cdot)&=\lambda_{jk}^S(\cdot)={p_{jk}} \lambda_j, \quad
\end{split}
\end{align}
\begin{equation}
\lambda_j =
\begin{cases}
\lambda_d \quad \text{if cluster $j$ is in the CBD} \\
\lambda_o \quad \text{if cluster $j$ is in the periphery} \\
\end{cases}       
\label{location_rate}
\end{equation}
where $\lambda_d$ and $\lambda_o$ are constant, and $p_{jk}$ is computed by decision rule (\ref{decision-rule}). Since the mobility rates for $j=1,2,...,J$ and $k\in N_G(j)$ are constant, the behavior of propagation process can be approximated by ordinary differential equations \eqref{eqn:diff_eqn} for the same reason as the previous model (Section~\ref{sssec:2.B.1}).
 
\subsubsection{Traffic density dependent mobility model}  
\label{sssec:2.B.3}
As in the previous two cases, the differential equations~\eqref{eqn:diff_eqn} can cater to traffic density dependent mobility model. Vehicle speed depends on the traffic density on the road with density being inversely related to speed; this is described by models such as the Greenshields model, the Drew model, and
the Pipes-Munjal model. The generalized form of these models~\citep{haefner1998traffic,kuhne1991macroscopic,wang2009speed} can be expressed as $v=v_f\left[ 1-\left({k}/{k_{jam}} \right)^a\right]^b$ \label{eqn:speed-density} where $v_f$ is free flow speed, $k_{jam}$ is jam density, and v and $k$ are speed and density respectively.

Motivated by these studies, we introduce the traffic density-dependent mobility rate $\lambda_{jk}(\cdot)$ from cluster $j$ to $k\in N_G(j)$, which depends on the relative density of both the cluster $j$ and $k$, where the relative density is defined as the fraction of vehicles located in the cluster $j$ and $k$. As the fraction of vehicles located in cluster $j$ and $k$ increases, the movement speed from cluster $j$ to $k$ is slowed down, which implies that the mobility rate from the origin cluster $j$ to the neighboring cluster $k$ decreases. Concretely, when the traffic density of the origin cluster $j$ is high, movement beyond this region is restricted, and if the traffic density of the destination cluster $k$ is high, it is also difficult to enter this region. The mobility rate reflecting this mobility characteristic is set to 
\begin{align}
\begin{split}
\lambda_{jk}^I(\cdot)&=\lambda_{jk}^S(\cdot)=\lambda \cdot {p_{jk}}  \cdot \left[1-\left(\sum_{i\in \{j,k\}}(I_i+S_i)\right)^a\right]^b 
\label{eqn:mob_den_avg_origin}
\end{split}
\end{align}
where $\lambda$ is constant, $p_{jk}$ is computed by decision rule (\ref{decision-rule}), and $a, b\geq 1$. By controlling parameters $a$ and $b$, we can reflect a general form of relationship between vehicles speed and traffic density. The mobility rate function $\lambda_{jk}$ is Lipschitz continuous on $E$ for $j=1,2,...,J$ and $k\in N_G(j)$ since the function is continuously differentiable on $E$. Thus, by results in the Appendix, the dynamics of the information propagation converge to the solution of the differential equation \eqref{eqn:diff_eqn}.

\section{Results of empirical validation} \label{sec:3}
We now empirically validate the mathematical model (Section~\ref{sec:2}) when the underline assumptions of the model are relaxed. The analytical result in the previous section ensures convergence of the propagation process results to solutions of differential equations only when the number of vehicles goes to infinity. To reflect real options, the number of vehicles must be finite. Therefore, we investigate the effect of a finite number of vehicles on various mobility and communication characteristics that exist in the real world. We first consider statistical models which we call synthetic models (Section~\ref{ssec:3.1}), and subsequently consider the actual vehicle trajectory data collected on different road topologies (Section~\ref{ssec:3.2}). In the latter case, the statistical communication process is superimposed on the actual trajectory data since there is no data currently available for the communication process. With a finite number of vehicles, we show that there is an excellent match between the result of the propagation process simulation and the model solution even for a moderate number of vehicles in the statistical models. We additionally show there is an acceptable match even for the actual trajectory data that does not satisfy the statistical assumption of exponential sojourn time under which convergence is guaranteed. Throughout this result section, the ordinary differential equations were solved using \texttt{ode} function from the \textbf{deSolve} package of R. 
\subsection{Synthetic models} \label{ssec:3.1}

We show that the output of the differential equations~\eqref{eqn:diff_eqn} matches simulations for different statistical mobility models with a finite number of vehicles. In Section~\ref{result:temporal}, we consider the mobility model of Section~\ref{sssec:2.B.1} which captures the temporal variation of the traffic density and routing. In Section~\ref{sssec:3.1.2}, we apply the location dependent mobility model of Section~\ref{sssec:2.B.2}. In Section~\ref{result:den_dep_avg}, we apply the traffic density dependent model of Section~\ref{sssec:2.B.3}. For the validation, we consider the grid road topology introduced in Section~\ref{ssec:2.2} (We will consider the other road topology in Section~\ref{ssec:3.2}). As shown in Figure~\ref{fig:grid_distance}, there are six avenues and six streets. The CBD is the shaded area in Figure~\ref{fig:grid_distance} and the rest is the periphery. All road segments are assumed to be two-way roads which are set to be composed of two clusters corresponding to the opposite directional roads. Since two clusters on the same road segment are sufficiently close to each other, the vehicles located in these can communicate. Therefore, the adjacency matrix of the communication network $G(V,E')$ is given by $a'_{ij} = a'_{ji} = 1$ if $i$ and $j$ are in the same road segment, otherwise $a'_{ij} = a'_{ji} = 0$. We set the communication parameter to $\beta_{ij} = 3$ if $a'_{ij}=1$ and $\beta_{ij} = 0$ otherwise. At initial time, $N$ vehicles are uniformly distributed in $J$ clusters, where $J=120$. Thus, each cluster has $n=N/J$ vehicles at initial time. The information of interest initially begins to propagate from 10\% of vehicles located in the lower left cluster (equivalently, 0.1n vehicles), which are located in peripheral areas. To study the degree of information propagation, we introduce $\rho (t)$, the fraction of overall vehicles that are informed at time $t$. 
\begin{figure}[!t]
	\centering
	\subfloat[Fraction of informed vehicles]{%
		\includegraphics[width=.45\linewidth]{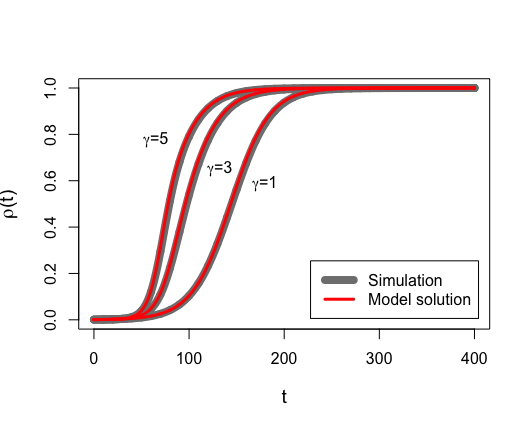}%
		\label{fig:temporal_result}
	}%
	\subfloat[Fraction of vehicles in CBD]{%
		\includegraphics[width=.45\linewidth]{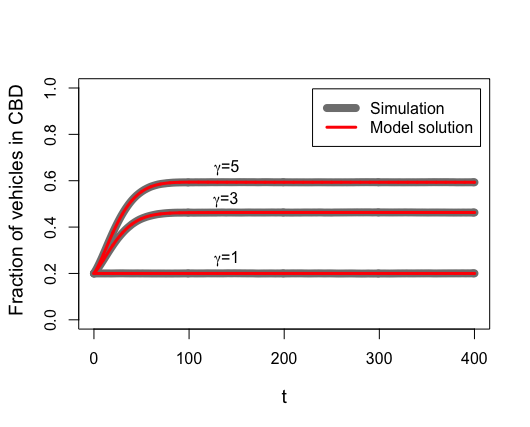}%
		\label{fig:Fraction of vehicles in CBD over time}
	}%
	\captionsetup{singlelinecheck = false, justification=raggedright, labelsep=space}
	\caption{The gray lines represent the average of 200 simulation runs of the information propagation, and the red lines are the solutions of the ordinary differential equations.} 
\end{figure}
\begin{figure}[!t]
	\centering
	\subfloat[Maximum deviation]{%
		\includegraphics[width=.45\linewidth]{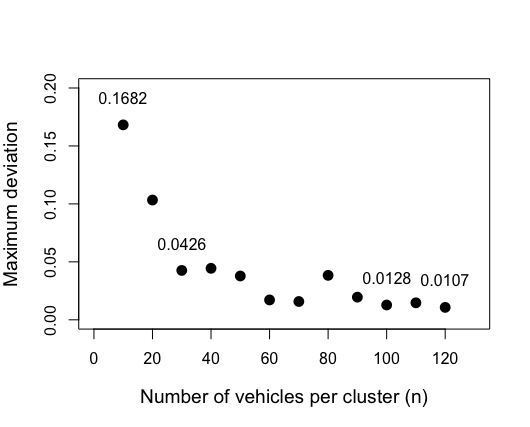}
		\label{fig:maxdev_num}
	}%
	\subfloat[$\rho (t)$ of changing the number of vehicles per cluster]{%
		\includegraphics[width=.45\linewidth]{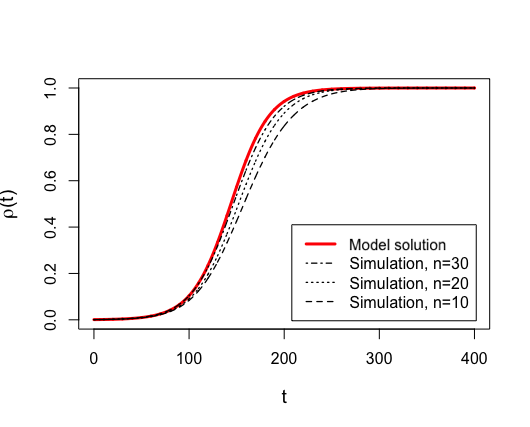}
		\label{fig:rho_num}
	}%
	\captionsetup{singlelinecheck = false, justification=raggedright, labelsep=space}
	\caption{Both (a) and (b) show that with the increase the number of vehicles per cluster, simulation results approach the solutions of the differential equations. We consider the average of 200 simulation runs in both cases. Even for $n=30$, the maximum deviation is as little as about 0.0426.}
\end{figure}

\begin{figure}[!t]
	\centering
	\subfloat{%
		\includegraphics[width=.19\linewidth]{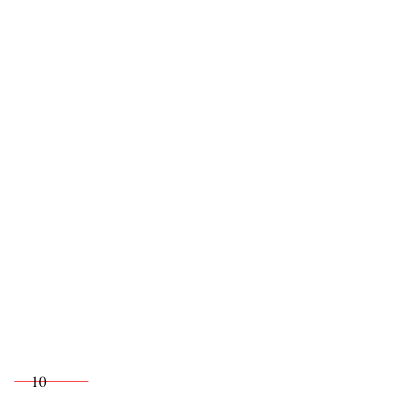}
	}%
	\hspace{5pt}
	\vrule
	\hspace{7pt}
	\subfloat{%
		\includegraphics[width=.19\linewidth]{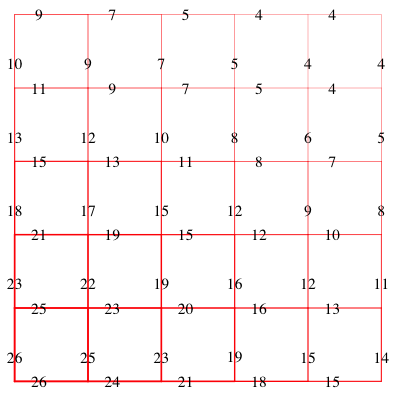}
	}%
	\subfloat{%
		\includegraphics[width=.19\linewidth]{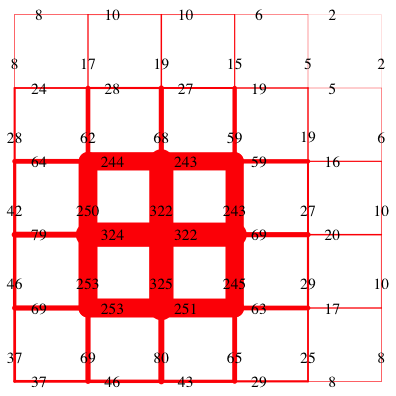}
	}%
	\subfloat{%
		\includegraphics[width=.19\linewidth]{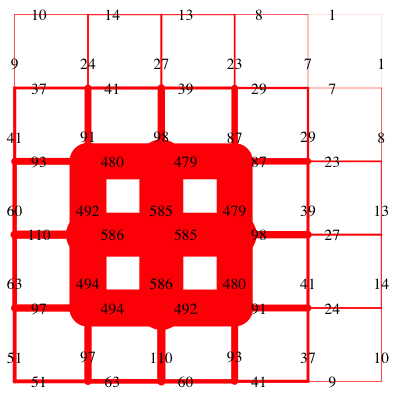}
	}%
	\\
	\vspace{-10pt}
	\addtocounter{subfigure}{-4}
	\subfloat[t=0s]{%
		\includegraphics[width=.19\linewidth]{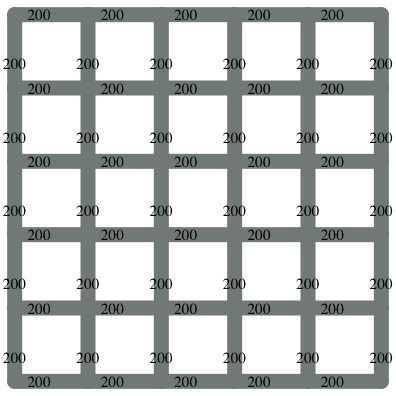}
	}%
	\hspace{5pt}
	\vrule
	\hspace{7pt}
	\subfloat[t=90s, $\gamma=1$]{%
		\includegraphics[width=.19\linewidth]{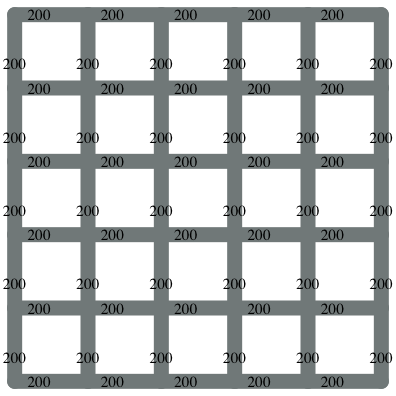}
	}%
	\subfloat[t=90s, $\gamma=3$]{%
		\includegraphics[width=.19\linewidth]{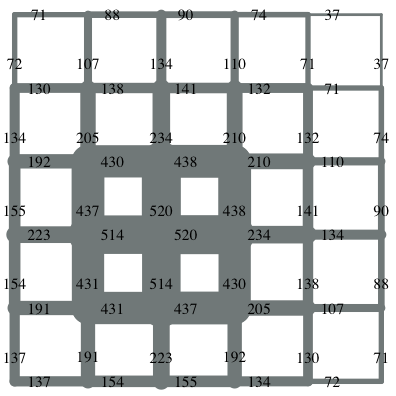}
	}%
	\subfloat[t=90s, $\gamma=5$]{%
		\includegraphics[width=.19\linewidth]{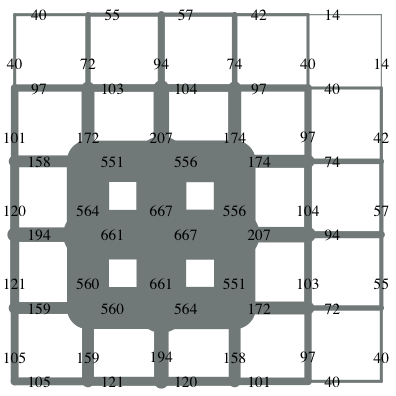}
	}%
	\captionsetup{singlelinecheck = false, justification=raggedright, labelsep=space}
	\caption{Geographical representation of traffic density and information propagation for various gammas under the same initial conditions with $n=100$ (i.e., 200 vehicles per road segment). The thickness of the road segment is linearly proportional to the number of vehicles, that is, when the thickness corresponding to one vehicle is $x$, the thickness of the vehicle $v$ is expressed by $v\cdot x$. The number written on each road segment indicates the number of vehicles. The upper row of red corresponds to the number of vehicles informed in each road segment, and the lower row of gray corresponds to the total number of vehicles in each road segment. The first column shows the initial distribution of vehicles, which is same for all $\gamma$. The remaining columns show the distribution of vehicles at time $t = 90s$ for different $\gamma$.} 
	\label{fig:gamma=1,3,5,ode}
\end{figure}
\subsubsection{Temporal variation of traffic density and routing} \label{result:temporal}
We show that the simulation of the propagation process with the mobility model in Section~\ref{sssec:2.B.1} closely matches the output of the differential equations~\eqref{eqn:diff_eqn} for a finite number of vehicles. We subsequently use the differential equations to understand the temporal and spatial propagation of the information. The mobility parameter $\lambda$ in \eqref{eqn:temporal_rate} is set to $\lambda = 0.1$. We set n=100, thus $N = n\cdot $J$ = 12000$. Under these settings, we compare the solutions of the corresponding differential equations with an average of 200 runs of the propagation process simulations.
Then, the results of the propagation process and the solution of the corresponding differential equations are compared. We consider different $\gamma$ values reflecting the temporal variation of traffic density and routing. Figure~\ref{fig:temporal_result} shows that the simulations of the propagation process closely match the solutions of the differential equations. The largest deviation between them is 0.0128 for $\gamma=1$, 0.0191 for $\gamma=3$, and 0.0226 for $\gamma=5$.

Now, we investigate the impact of a relatively small number of vehicles on our model. For non-rush hour ($\gamma = 1$), Figure~\ref{fig:maxdev_num} shows that the greater the number of cars, the smaller the maximum deviation between the two results. Even for a small $n$, the solution of the differential equations well approximates the simulation of the propagation process: the maximum deviation between the two is (a) 0.0426 for $n=30$ and (b) 0.1682 for $n=10$. This means that asymptotic results are obtained even for a very small number in practice. 
\begin{figure}[!b]
	\centering
	\subfloat[Fraction of informed vehicles]{%
		\includegraphics[width=.45\linewidth]{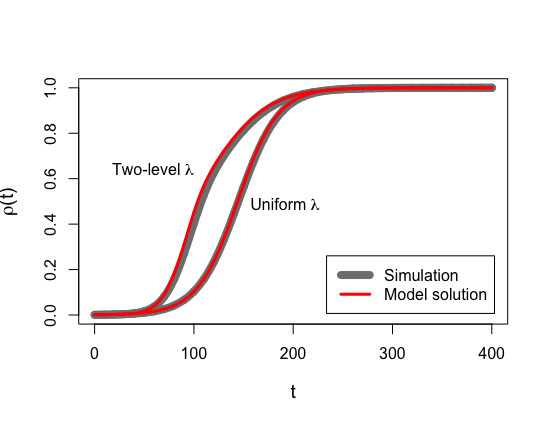}
		\label{fig:location_inf_rat_a}
	}%
	\subfloat[Fraction of vehicles in CBD]{%
		\includegraphics[width=.46\linewidth]{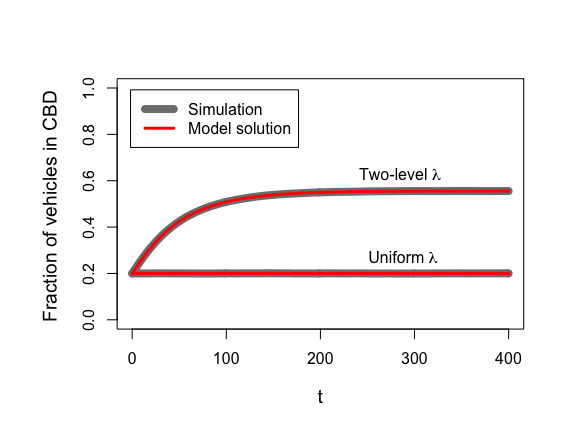}
		\label{fig:location_inf_rat_b}
	}%
	\captionsetup{singlelinecheck = false, justification=raggedright, labelsep=space}
	\caption{The curves of the two-level $\lambda$ are the result of applying a different $\lambda$ value depending on whether the region is CBD or peripheral region, and the uniform $\lambda$ case is the result of applying the same $\lambda$ regardless of region. The simulation results are averaged over 200 runs.}
	\label{fig:location_inf_rat}
\end{figure}

Now that we have verified that the differential equations~\eqref{eqn:diff_eqn} can capture information propagation reasonably accurately even with a moderate number of vehicles, we now use them to understand the characteristics of information propagation. In the case of non-rush hour ($\gamma=1$), as shown in Figure~\ref{fig:Fraction of vehicles in CBD over time}, the information slowly spreads over the entire area without an upsurge in particular areas. 
On the other hand, in the case of morning rush hour ($\gamma > 1$), there is a significant increase in the number of vehicles in the CBD, and as a result, information spreads very quickly. The larger the $\gamma$, the higher the traffic density in the CBD; thus, the information spreads quicker in the CBD than in the periphery, as shown in the geographical representation (Figure~\ref{fig:gamma=1,3,5,ode}).
\subsubsection{Location dependent mobility rate} \label{sssec:3.1.2}

\begin{figure}[!t]
	\centering
	\subfloat[]{%
		\includegraphics[width=.34\linewidth]{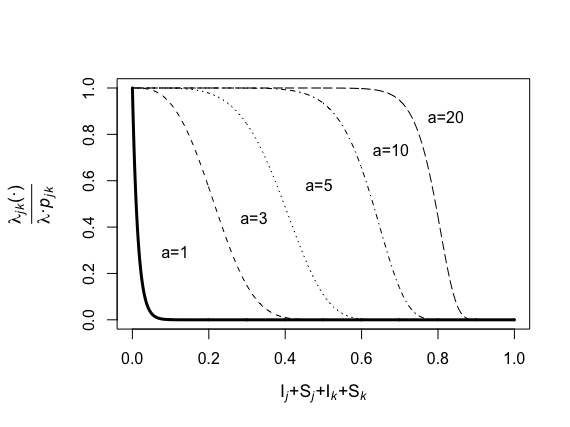}
		\label{fig:mob_den_avg_origin_g}
	}%
	\subfloat[]{%
		\includegraphics[width=.32\linewidth]{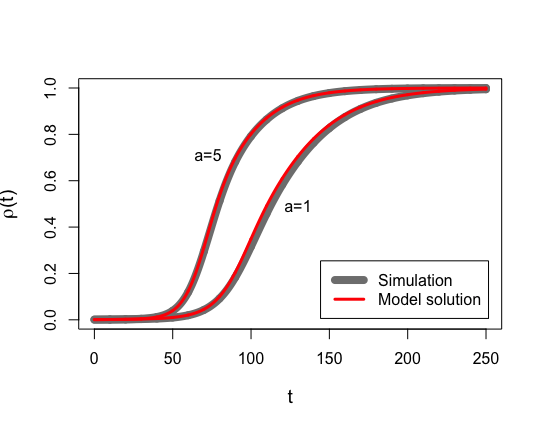}
		\label{fig:mob_den_avg_b}
	}%
	\subfloat[]{%
		\includegraphics[width=.33\linewidth]{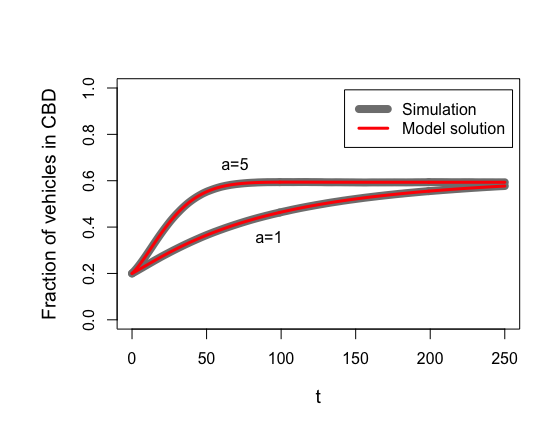}
		\label{fig:mob_den_avg_c}
	}%
	\caption{(a) Relation between mobility rate devided by constant factor ($\frac{\lambda_{jk}(\cdot)}{\lambda\cdot p_{jk}}$) and traffic density $(I_j+S_j+I_k+S_k)$, for a given origin cluster $j$ and destination cluster $k$. (Recall that the mobility rate in Section~\ref{sssec:2.B.3} is given by $\lambda_{jk}(\cdot)=\lambda\cdot p_{jk}\cdot\left[1-(I_j+S_j+I_k+S_k)^a\right]^b$) (b) Fraction of informed vehicles over time. (c) Fraction of vehicles located in CBD over time. For both (b) and (c), the gray lines are simulation results of averaging 200 simulation runs, and the red lines are the solutions of the approximate ordinary differential equations. For all three figure, b=70.} 
\end{figure}
\begin{figure}[!t]
	\centering
	\subfloat{%
		\includegraphics[width=.18\linewidth]{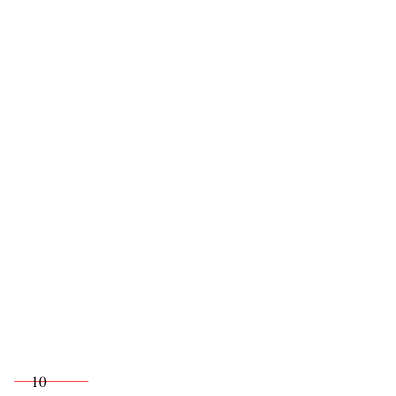}
	}%
	\vrule
	\hspace{.5pt}
	\subfloat{%
		\includegraphics[width=.18\linewidth]{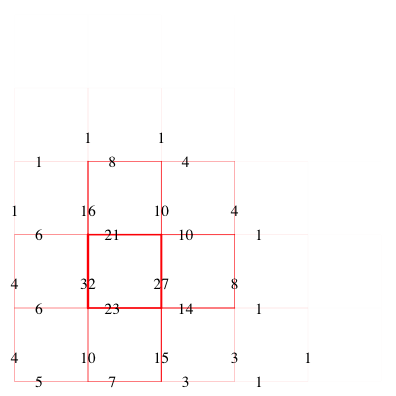}
	}%
	\subfloat{%
		\includegraphics[width=.18\linewidth]{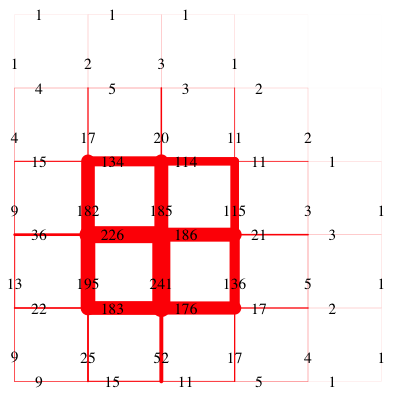}
	}%
	\vrule
	\hspace{.5pt}
	\subfloat{%
		\includegraphics[width=.18\linewidth]{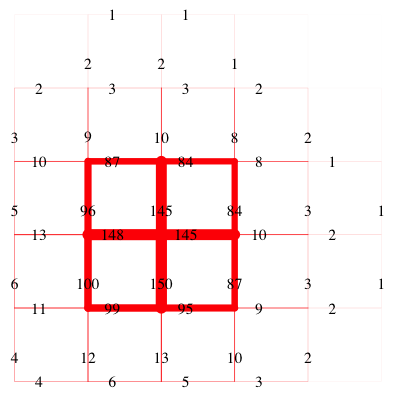}
	}%
	\subfloat{%
		\includegraphics[width=.18\linewidth]{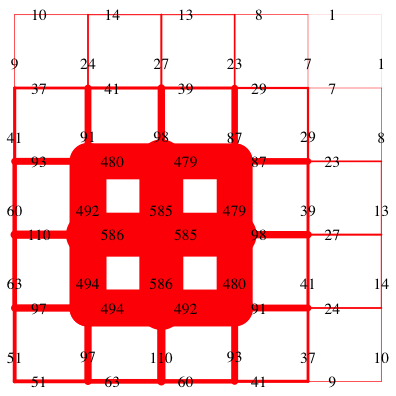}
	}%
	\\
	\vspace{-10pt}
	\addtocounter{subfigure}{-5}
	\subfloat[t=0s]{%
		\includegraphics[width=.18\linewidth]{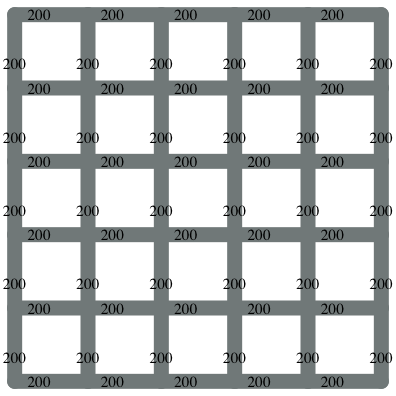}
	}%
	\vrule
	\hspace{.5pt}
	\subfloat[t=60s, a=1]{%
		\includegraphics[width=.18\linewidth]{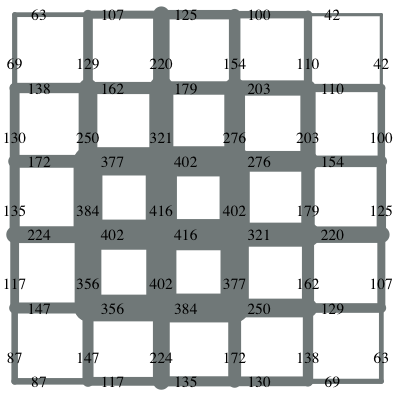}
	}%
	\subfloat[t=90s, a=1]{%
		\includegraphics[width=.18\linewidth]{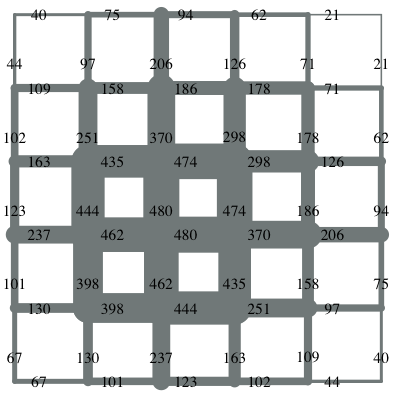}
	}%
	\vrule
	\hspace{.5pt}
	\subfloat[t=60s, a=5]{%
		\includegraphics[width=.18\linewidth]{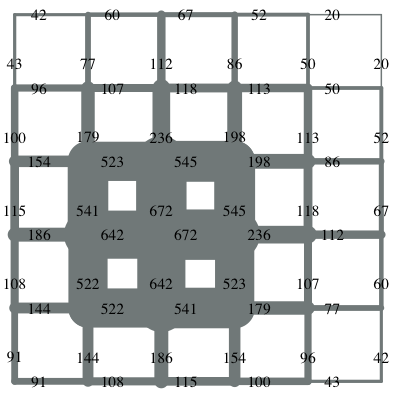}
	}%
	\subfloat[t=90s, a=5]{%
		\includegraphics[width=.18\linewidth]{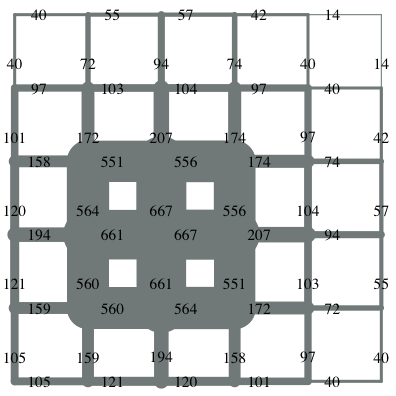}
	}%
	\captionsetup{singlelinecheck = false, justification=raggedright, labelsep=space}
	\caption{The upper row of red corresponds to the number of informed vehicles in each cluster, and the lower row of gray corresponds to the number of vehicles in each cluster. The thickness of the road segment is linearly proportional to the number of vehicles, that is, when the thickness corresponding to one vehicle is $x$, the thickness of the vehicle $v$ is expressed by $v\cdot x$. The first column represents the initial distribution of vehicles, which is the same for all values of $a$. The remaining columns show the times and the values of $a$ as indicated.}
	\label{fig:den_dep_avg_geo}
\end{figure}

We now consider the location dependent mobility model introduced in Section~\ref{sssec:2.B.1}. In this subsection, two different mobility rate values are applied to roads in the CBD and the periphery respectively, which we call \textit{two-level mobility rate}. We set the mobility rate of the CBD to a lower value, reflecting a lower speed limit in the CBD. For the two-level mobility rate case, we choose $\lambda_d = 0.05$ for the CBD and $\lambda_o = 0.1$ for the periphery in (\ref{location_rate}); for the uniform mobility rate case, we choose $\lambda_d = \lambda_o = 0.1$ for both the CBD and the periphery. The number of vehicles per cluster is set to $n = 100$ at $t=0$. Also, $\gamma$ is assumed to be 1 (non-rush hour). Figure~\ref{fig:location_inf_rat_a} shows that the propagation process for this two-level mobility rate is also well approximated by solutions of the corresponding differential equations~\eqref{eqn:diff_eqn}: The maximum deviation between them is only 0.0484. We now compare the propagation of information under the two-level mobility rate with that for uniform mobility rate. Since the mobility rate in the CBD is lower than in the periphery, the average speed of vehicles in the CBD is slower. Therefore, even in the case of non-rush hour with $\gamma=1$, once vehicles enter the CBD, the vehicles in the CBD take longer to move, resulting in the concentration of vehicles in the CBD (Figure~\ref{fig:location_inf_rat_b}). Concentrated traffic leads to faster information propagation as shown in Figure~\ref{fig:location_inf_rat_a}. 

\subsubsection{Traffic density dependent mobility model}   \label{result:den_dep_avg}
Next we consider the mobility model of Section~\ref{sssec:2.B.3} where the mobility rate depends on the traffic density of both the origin and destination cluster. we set the mobility rate to depend on the traffic density of both the origin and destination clusters. For \eqref{eqn:mob_den_avg_origin}, we choose $\lambda = 0.1$. Recall that the sensitivity of the mobility rate to traffic density is determined by parameters $a$ and $b$ of \eqref{eqn:mob_den_avg_origin}. As Figure~\ref{fig:mob_den_avg_origin_g} shows for fixed $b$, the mobility rate $\lambda_{jk}$ decreases with traffic density of cluster $j$ and $k$; this is sharp for relatively small $a$ and more gradual as $a$ increases. To illustrate, we fix the parameter $b=70$ and compare the results for $a=1$ and $a=5$. In both cases, we consider the movement pattern of the morning rush hour ($\gamma=5$). As Figure~\ref{fig:mob_den_avg_b} shows, the differential equations closely approximate the propagation process; The maximum deviation between them is 0.0275 at $a = 1$ and 0.0187 at $a = 5$.

Now using the geographical representation of information propagation in Figure~\ref{fig:den_dep_avg_geo}, we investigate how information propagates for traffic density dependent model. Since we are considering morning rush hour, the vehicles congregate in the CBD over time, thus information propagates faster therein. This phenomenon is pronounced for larger values of $a$ as the mobility rate increases with $a$ given fixed $b$ and traffic density as can be seen in Figure~\ref{fig:mob_den_avg_origin_g}; this is supported by the results in Figure~\ref{fig:location_inf_rat_b}.

\subsection{Empirical validation with traffic trace data} \label{ssec:3.2}
When the mobility process is exponential, even for a finite number of vehicles, the solution of the differential equations closely approximates the dynamics of the propagation process (Section~\ref{ssec:3.1}). In this section, we show that the temporal and spatial flow of information is well approximated by the solution of differential equations even when the mobility process is not exponential, and the number of vehicles $N$ is finite. Towards that end, we use we use microscopic vehicle trajectory data collected to validate our model empirically. Thus far, we have only considered grid topology, but we will also consider other topologies in this section. In this case, we show that the output of the differential equations reasonably matches the dynamics of information propagation for various road topologies.

The three vehicle trajectory data we use are collected from different road topologies of varying complexity, unlike the grid topologies considered earlier; U.S. Highway 101 in Los Angeles, California (Section~\ref{ssec:4.1}), Peachtree Street in Atlanta, Georgia (Section~\ref{ssec:4.2}), and Europarc Roundabout in Creteil, France (Section~\ref{ssec:4.3}). We manually divide each topology into $J$ clusters, and define the mobility network $G$ and the communication network $G'$. The directed mobility network $G$, which is represented as a directed edge between clusters, is determined by the existence of a trajectory in which the vehicle moves from one cluster to another. To indicate that vehicle-to-vehicle communication in the cluster is possible, the diagonal element of the adjacency matrix corresponding to the communication network is set to 1 as a baseline for all three data sets. In addition, if the distance between neighboring clusters is close enough, the corresponding element is set to 1 to enable inter-cluster communication according to the characteristics of the road.
The mobility rate between clusters is extracted from the vehicle trajectory data, and is applied to the ordinary differential equation of our model to estimate the information propagation. We superimpose the statistical communication process on the actual trajectory data, and compare the result with the solution of the differential equations. As a result, not only does this show that our model is applicable to arbitrary road topologies, but it also shows that simulation results using even actual trajectory data are well approximated by model solutions.
To extract the mobility rates from the data, we will first classify the clusters into three categories. Let $S$ be the set of clusters of the study area where information propagation occurs. Note that the mathematical model considers the fixed set of vehicles in the system. In the real transportation network, there would be entrances from outside and also exit to outside. To incorporate the impact of the entrances and exits, we introduce a set of virtual clusters $A$ and $B$ where $A$ and $B$ are the respective set of clusters corresponding to the entry and exit roads respectively entering and leaving the study area. Let $O=S\cup A\cup B$. Here we consider that the mobility rate of a vehicle does not depend on whether it is informed or non-informed.

We now describe how we obtain mobility rate $\lambda_{i,j}$ of the analytical model from the trajectory data ($\lambda_{i,j}$ moving from cluster $i\in S$ to $j\in O$, such that $i\neq j$). First, from the trajectory data, we compute $\lambda_i$, which is the reciprocal of the average time of staying in cluster $i$. Then we compute the fraction of vehicles that move to cluster $j$ among the vehicles located in cluster $i$, and denote this as $p_{i,j}$. By multiplying $\lambda_i$ by probability $p_{i,j}$, the mobility rate from cluster $i$ to $j$, $\lambda_{i,j}$, can be computed. If the origin cluster of the movement is the entry road, the mobility rate $\lambda_{i,j}$ moving from cluster $i\in A$ to $j\in S$, such that $i\neq j$, is given by $\widetilde{\lambda}_{i,j}/N$ where $N$ is the total number of vehicles and $\widetilde{\lambda}_{i,j}$ denotes the number of vehicles that enter cluster $j\in S$ from cluster $i\in A$ per unit time.

\subsubsection{U.S. Highway 101 in Los Angeles, California} \label{ssec:4.1}
\begin{figure}[pht]
	\centering
	\subfloat[]{%
		\includegraphics[width=.37\linewidth]{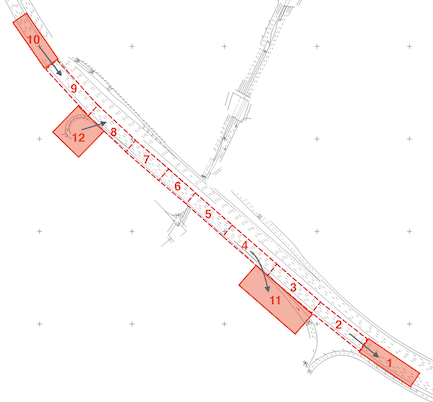}
		\label{fig:us101a}
	}%
	\hspace{10pt}
	\subfloat[]{%
		\includegraphics[width=.45\linewidth]{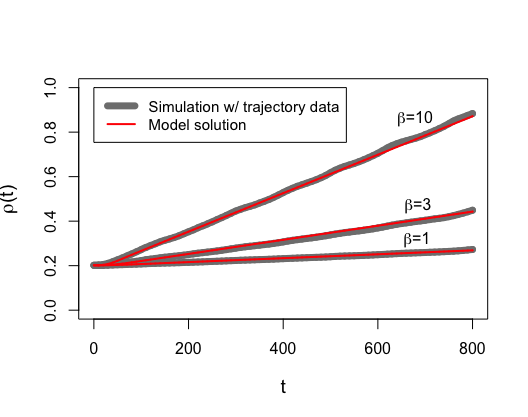}
		\label{fig:us101b}
	}%
	\captionsetup{singlelinecheck = false, justification=raggedright, labelsep=space}
	\caption{(a) U.S highway 101 broken into clusters. The vehicles enter the study area from the cluster 10 or 12 and leaves the study area toward the cluster 1 or 11. (b) Fraction of informed vehicles over time. The information spreading simulation based on the actual trajectory data is well approximated by the theoretical predictions from the ordinary differential equations~\eqref{eqn:diff_eqn}. The gray lines are simulation results of averaging 30 simulation runs for each $\beta$, and the red lines are the solutions of the differential equations.}
\end{figure}
We use the microscopic vehicle trajectory data of the southbound U.S. highway 101 in Los Angeles, California, that had been collected under the auspices of the Next Generation Simulation (NGSIM) program~\citep{ngsim}. This data includes geographical location information for each vehicle on the southbound U.S. highway 101 of 600 meters in length. We consider the total number of vehicles that exist in the system at any point during the observation period ($N=1993$). As shown in Figure \ref{fig:us101a}, we divide the road into $J=12$ clusters. The study area in which we conducted information propagation studies is a set of clusters $S=\{2,3,...,9\}$, which is depicted as a dotted rectangle in Figure \ref{fig:us101a}. Vehicles enter the system through clusters $\{8,9\}$, and leaves the system from clusters $\{2,4\}$. We introduce a set of virtual entry clusters and exit clusters as shown in the Figure~\ref{fig:us101a}; thus $A=\{10,12\}$ and $B=\{1,11\}$. The set of clusters $A$ and $B$ corresponding to the entry and exit roads respectively are represented by shaded solid line rectangles. The actual trajectory data for the first 831.7 seconds out of the entire data was used for this study, and all vehicles entering the study area were regarded as separate vehicles. There is no vehicle in the study area at an initial time, and we assumed that approximately 20\% of all incoming vehicles, that is, 402 out of 1993 incoming vehicles, had already received the information before they entered the study area. The adjacency matrix for the communication network is set to $a'_{ij} = 1$ if $i \in S$ and $i = j$, otherwise $a'_{ij} = 0$. As shown in Figure \ref{fig:us101b}, there is an excellent match between the simulation result using the actual trajectory data and the solutions of corresponding differential equations. The maximum deviations between the average of 30 simulation runs and the solution of the differential equation are 0.0039 for $\beta=1$, 0.0112 for $\beta=3$, and 0.0151 for $\beta=10$.

\subsubsection{Peachtree Street in Atlanta, Georgia} \label{ssec:4.2}
\begin{figure}[htp]
	\centering
	\subfloat[]{%
		\includegraphics[width=.18\linewidth,height=220pt]{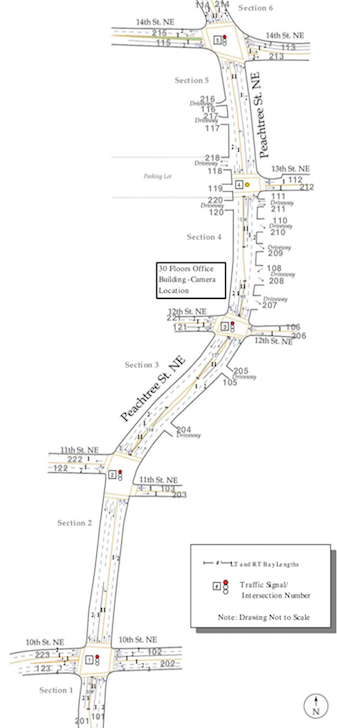}
		\label{fig:peachtree_a}
	}%
	\subfloat[]{%
		\includegraphics[width=.18\linewidth,height=220pt]{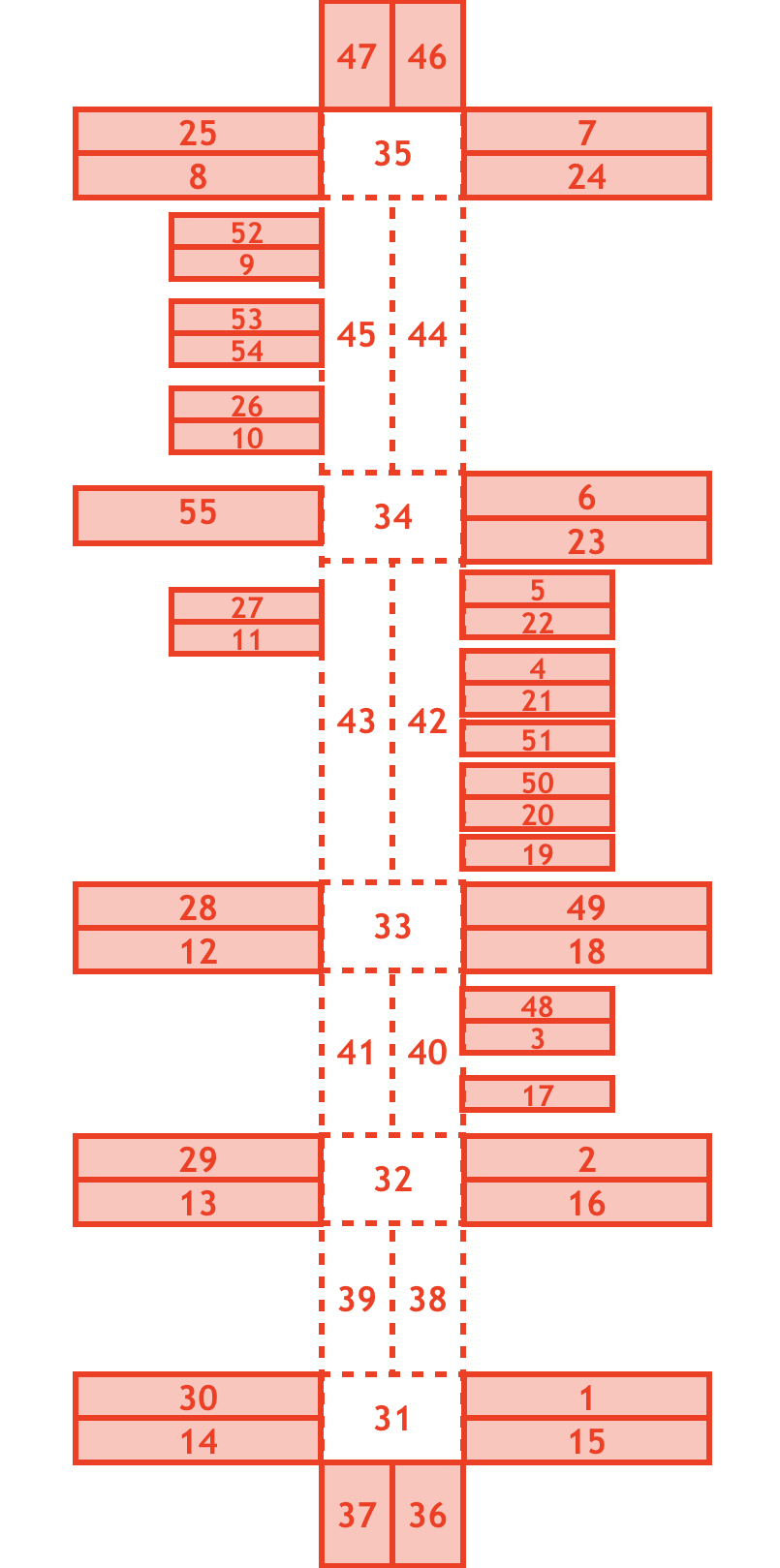}
		\label{fig:peachtree_b}
	}%
	\hspace{10pt}
	\subfloat[]{%
		\includegraphics[width=.47\linewidth]{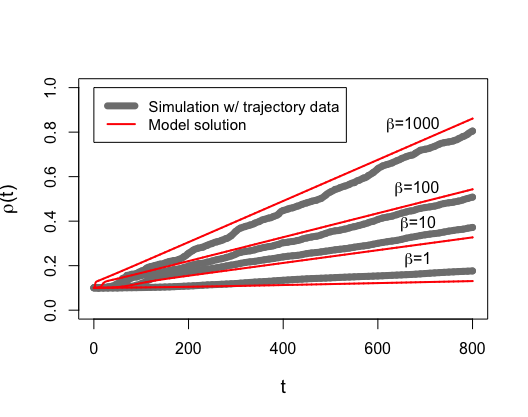}
		\label{fig:peachtree_c}
	}%
	\captionsetup{singlelinecheck = false, justification=raggedright, labelsep=space}
	\caption{(a) Peachtree street schematic~\citep{ngsim}. (b) Clustered Peachtree street (c) Fraction of informed vehicles over time. The information spreading simulation based on the actual trajectory data is well approximated by the theoretical predictions from the ordinary differential equations~\eqref{eqn:diff_eqn}. The gray lines are simulation results of averaging 30 simulation runs for each $\beta$, and the red lines are the solutions of the differential equations.}
\end{figure}
We use the microscopic vehicle trajectory data of the Peachtree street in Atlanta, Georgia, that had been also collected under the auspices of the Next Generation Simulation (NGSIM) program~\citep{ngsim}. This data includes geographical location information for each vehicle on the two-way street of 640 meters in length with 5 intersections, which are more complex than the previous one-way road topology. We consider the total number of vehicles that exist in the system at any point during observation period (N=2298). As shown in Figure \ref{fig:peachtree_b}, we divide the road into $J=55$ clusters. The actual trajectory data for the first 1044.2 seconds was used for this study, and all vehicles entering the study area were considered separate vehicles\footnote{In the original dataset~\citep{ngsim}, the \texttt{Total\_Frames} attribute represents a total number of frames in which the vehicle appears in the system. The \texttt{O\_Zone} and \texttt{D\_Zone} attributes represent the place where the vehicles enter and exit the system respectively. Records with the same value of \texttt{Vehicle\_ID} but different values of \texttt{Total\_Frames}, \texttt{O\_Zone}, and \texttt{D\_Zone} attributes show that the same vehicle has entered the system multiple times. In this study, every time a vehicle enters the system, we have considered it a new vehicle.}. In addition, we clean up the data to exclude obvious instances of data error and vehicles which do not enter the study area; we consequentially use 98.4\% of the original number of vehicles. As shown in Figure \ref{fig:peachtree_b}, the study area in which we conducted information propagation studies is a set of clusters $S=\{31,32,...,35,38,39,...,45\}$. Vehicles enter the study area from the virtual clusters $A=\{1,2,...,14,36,47,49,50,54,55\}$, and leaves the study area through the virtual clusters $B=\{15,16,...,30,37,46,48,51,52,53\}$. We consider that the vehicles in the study area are not informed at an initial time, but we assumed that 30\% of vehicles entering the study area from the cluster $36\in A$ and $47\in A$, that is, 231 out of 770 incoming vehicles, had received the information before they enter the study area. The adjacency matrix for the communication network is set to $a'_{ij} = 1$ if $i \in S$ and $i = j$. In addition, given their proximity, $a'_{38,39}=a'_{39,38}$, $a'_{40,41}=a'_{41,40}$, $a'_{42,43}=a'_{43,42}$, and $a'_{44,45}=a'_{45,44}$ are also set to 1, and all other elements are set to zero. Under these conditions, the information propagation simulation using the actual trajectory data on Peachtree street is well approximated by the model solution as shown in Figure \ref{fig:peachtree_c}. The maximum deviations between the average of 30 simulation runs and the solution of the differential equations are 0.0456 for $\beta=1$, 0.0448 for $\beta=10$, 0.0376 for $\beta=100$, and 0.0647 for $\beta=1000$.

\subsubsection{Europarc Roundabout in Creteil, France} \label{ssec:4.3}
\begin{figure}[htp]
	\centering
	\subfloat[]{%
		\includegraphics[width=.35\linewidth]{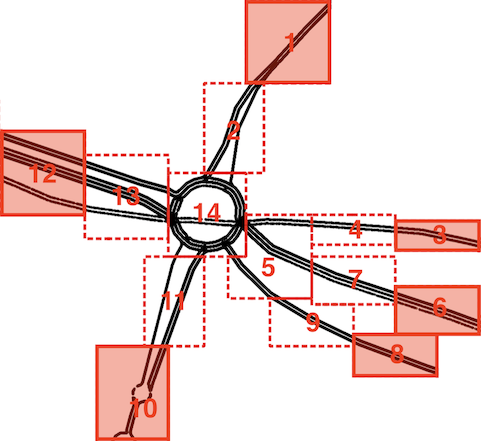}
		\label{fig3.21a}
	}%
	\hspace{15pt}
	\subfloat[]{%
		\includegraphics[width=.45\linewidth]{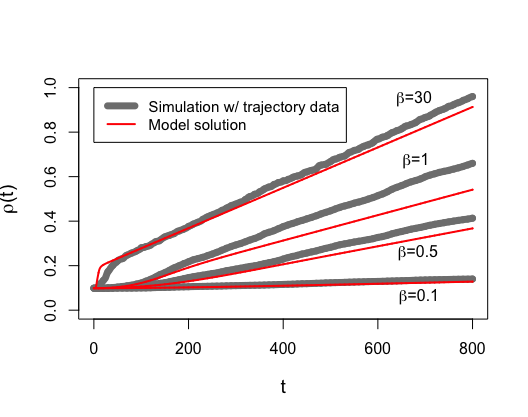}
		\label{fig3.21b}
	}%
	\captionsetup{singlelinecheck = false, justification=raggedright, labelsep=space}
	\caption{(a) Clustered Europarc roundabout road. (b) Fraction of informed vehicles over time. The information spreading simulation based on the actual trajectory data is well approximated by the theoretical predictions from the ordinary differential equations~\eqref{eqn:diff_eqn}. The gray lines are simulation results of averaging 30 simulation runs for each $\beta$, and the red lines are the solutions of the differential equations.}
\end{figure}
We use the Microscopic vehicle trajectory data of the Europarc Roundabout in Creteil, France, that had been generated by~\cite{LebMA:OnIRDMUVMT}. According to~\cite{LebMA:OnIRDMUVMT}, traffic information for Europarc roundabout was based on actual observations of vehicle flow, and manual counting were performed to generate origin/destination (O/D) matrix. This O/D matrix faithfully mimics the daily movement of the vehicle, from which a realistic synthetic data set of vehicle mobility is presented by~\cite{LebMA:OnIRDMUVMT}. We choose 800 seconds from the morning traffic data (7.15 AM to 9.15 AM) which have the peak number of vehicles (the peak occurs between time steps 4200 to 5000). This topology is much more complex than the previous two; specifically, this is not a grid topology, as it includes a roundabout road and 15 traffic lights. As shown in Figure~\ref{fig3.21a}, we divide the road into $J=14$ clusters. The study area in which we conducted information propagation studies is a set of clusters $S=\{2,4,5,7,9,11,13,14\}$. Vehicles enter the road from the set of virtual clusters $A=\{1,3,6,10,12\}$, and leave the study area toward the set of virtual clusters $B=\{1,3,10,12\}$.
Suppose that all vehicles entering the study area were regarded as separate vehicles. We consider that the vehicles in the study area are not informed at an initial time, but we assumed that approximately 11\% of vehicles entering the study area from the set of clusters $A$, that is, 66 out of 591 incoming vehicles, had received the information before they enter the study area. 
The adjacency matrix for the communication network is set to $a'_{ij} = 1$ if $i \in S$ and $i = j$, otherwise $a'_{ij} = 0$. 

As shown in Figure 12.b, there is a reasonable match between the solution of the differential equations~\eqref{eqn:diff_eqn}. The maximum deviation between the average of 30 simulation runs and the solution of the differential equation is 0.0124 for $\beta=0.1$, 0.0507 for $\beta=0.5$, 0.1185 for $\beta=1$, and 0.0741 for $\beta=30$. However, the match is not as close as in the previous cases. We now explain the reasons for this discrepancy. First, the total number of vehicles in this data is fairly low compared to the previous two datasets. In addition, the traffic movement becomes pulsed because of the presence of 15 traffic lights in a segment of the topology. This results in traffic synchronization which causes significant divergence between the mobility of this trajectory and the exponential mobility process; recall that the mathematical framework in Section~\ref{sec:2} has assumed the exponential mobility process.

\section{Use of the model} \label{sec:4}
We have shown an excellent match between the solution of the approximated differential equation and the propagation process results for a finite number of vehicles in the statistical models that reflect diverse realistic mobility and communication characteristics. We also showed that there is a fairly good match between the two results, even for actual trajectory data that does not satisfy the statistical assumptions under which convergence is guaranteed. Now, we will explain how these differential equation based models for information propagation can be utilized through concrete real world examples. As in the result section on the synthetic model, we assume a grid topology consisting of two-way roads, but we consider a larger city consisting of nine avenues and nine streets with the total number of clusters $J=288$. Likewise, the adjacency matrix of the communication network $G(V,E')$ is given by $a'_{ij} = a'_{ji} = 1$ if $i$ and $j$ are in the same road segment, otherwise $a'_{ij} = a'_{ji} = 0$. We apply the model with temporal variation of the traffic density and routing (Section)~\ref{sssec:2.B.1}. The communication parameter is set to $\beta_{ij} = \beta$ if $a'_{ij}=1$ and $\beta_{ij} = 0$ otherwise where $\beta$ is constant. Since we have verified the validity of the approximation by differential equations, we now use differential equations to understand information propagation characteristics in this section.

\subsection{Unexpected events} \label{unexpected_events}
\begin{figure}[htp]
	\centering
	\subfloat[Location of the event]{%
		\includegraphics[width=.3\linewidth]{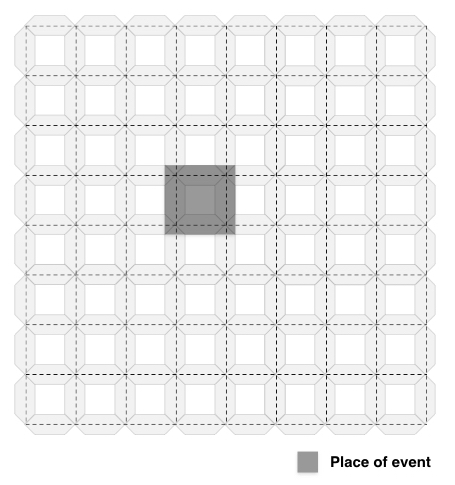}
		\label{arena_map}
	}%
	\hspace{30pt}
	\subfloat[Fraction of clusters]{%
		\includegraphics[width=.46\linewidth]{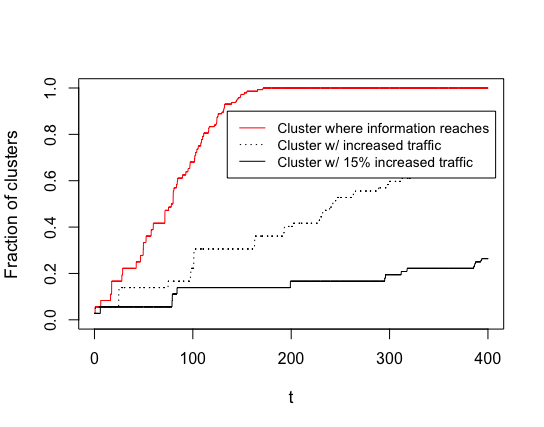}
		\label{arena_b}
	}%
	\captionsetup{singlelinecheck = false, justification=raggedright, labelsep=space}
	\caption{(a) Clustered grid road topology. Location of the event is represented by a dark shaded square. (b) The red solid line represents the percentage of clusters in which information arrives, the solid black line represents the percentage of congested clusters that have increased by more than 15\% over the initial traffic volume. The black dashed line represents the fraction of clusters with increased traffic (i.e. the fraction of clusters increases by more than one from the initial number of vehicles).} 
\end{figure}

Traffic conditions are known to vary widely and sometimes abruptly in a variety of unexpected events. For example, at the end of a major sporting event, unexpected victories can cause sudden increases in traffic in areas of celebration. Also, personal unscheduled visits by high-ranking officials (e.g., president) and celebrities can significantly increase traffic volume in certain areas. Gradually, the congestion from these hot spots spread all over the city including the periphery. This congestion can be ameliorated through the use of V2V communications. Vehicles parked near an event have information about the traffic congestion around the location of the event; these vehicles then transmit traffic congestion information to other vehicles through V2V communications while moving. Vehicles located in the periphery that have received the information can bypass congested roads, which can mitigate traffic congestion. Congestion control is particularly critical for vehicles in ride-sharing services in that it saves participants’ time and money. Besides, shared ride services can enhance societal utility by helping in the reduction of traffic congestion and thereby improving the travel experience of the drivers who are not involved in shared mobility. The first step toward this end is to assess the efficacy of V2V technology in mitigating traffic congestion due to sudden events by studying how quickly V2V can help spread information about congestion.

We model an unexpected event occurring in a central area of a city with a grid topology (Figure~\ref{arena_map}). It is also assumed that 12.5\% of the total number of vehicles $N = 14400$ are located on 4 road segments (equivalently, 8 clusters) surrounding the event site, and 45 vehicles are uniformly distributed in all other clusters. We set the parameters to $\beta = 10$, $\lambda = 0.05$, and $\gamma = 1$ (non-rush hour). Using our model, we can estimate how the traffic changes over time due to the events, and investigate how the information about the traffic propagate during the dispersion of the gathered vehicles. Our numerical computations reveal that information about the expected traffic congestion propagates faster than the spread of traffic congestion itself. As can be seen in Figure~\ref{arena_b}, the number of clusters in which information arrives (i.e., the number of clusters in which more than one vehicle receives information) increases much faster than the number of clusters with more than 15\% increase in traffic volume. These results have been demonstrated when the unexpected event occurs in the center of the grid. Because unexpected events can occur anywhere in a city, we investigate the impact of the location of the initial informed vehicles on information propagation in Section~\ref{initial_location}.

\subsection{Obstructions} \label{roadblocks}
\begin{figure}[htp]
	\centering
	\subfloat[Blocked road]{%
		\includegraphics[width=.3\linewidth]{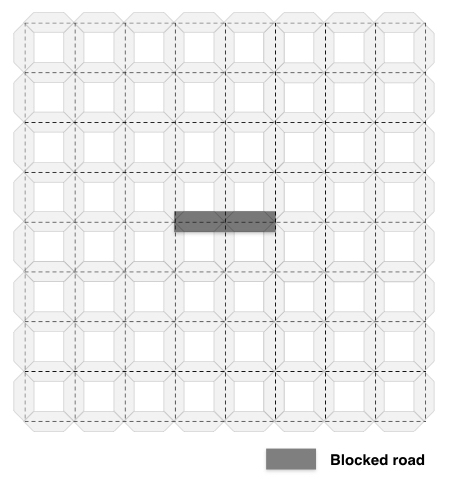}
		\label{fig:roadblocks_map_roadblcks2}
	}%
	\hspace{30pt}
	\subfloat[Fraction of clusters]{%
		\includegraphics[width=.43\linewidth]{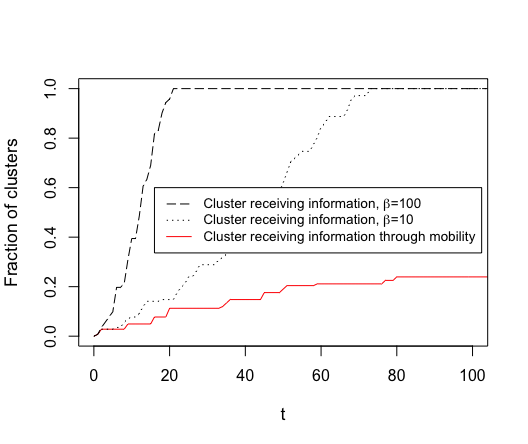}
		\label{fig:roadblocks2_result}
	}%
	\captionsetup{singlelinecheck = false, justification=raggedright, labelsep=space}
	\caption{(a) Clustered grid road topology. Blocked road segments are represented by dark shaded squares. (b) The red solid line represents the fraction of clusters in which one or more informed vehicles reach when there is no communication ($\beta=0$). This represents the rate at which information propagates solely due to mobility. The black dotted and black solid lines represent the fraction of clusters that one or more informed vehicles reach, in the presence of both communication and mobility, where $\beta=10$ and $\beta=100$, respectively. This represents the rate at which information propagates due to the combination of communication and mobility.}
	\label{fig:roadblocks_result}
\end{figure}
In many cases, roads are obstructed for reasons such as traffic accidents or road maintenance. Vehicles upstream of the obstruction must be detoured while vehicles that are in the queue immediately upstream of the obstruction must be discharged. There are several obvious advantages associated with dispersing the location information of obstructions as early as possible. Drivers who are heading for the obstruction can choose an alternate route even from a distance, reducing their own inconveniences as well as traffic congestion around the area. Towards that end, we assess how quickly the location information of the obstruction can be spread through the utilization of V2V technology.

We assume that two road segments in a central area of the city are blocked (Figure~\ref{fig:roadblocks_map_roadblcks2}). Information is propagated from queued vehicles, stopped from the obstruction. Let $R$ be the set of clusters that are blocked. If $j\in R$, $a_{ij}=0$ in the adjacency matrix of the mobility network, reflecting that the vehicle cannot enter the obstructed clusters. The mobility parameter $\lambda$ is set to $\lambda = 0.05$, and communication parameter $\beta$ is set to $\beta=10$ and $\beta=100$. The total number of vehicles $N=14400$ is uniformly distributed in the total number of clusters $J = 288$ at initial time, thus the initial number of vehicles per cluster is set to $n = 50$.
In Figure~\ref{fig:roadblocks_result}, we plot as a function of time the number of clusters in which (a) vehicles containing information on the obstruction location reach, and (b) the initially informed vehicles located on the obstructed road reach. Figure~\ref{fig:roadblocks_result} reveals that the result of the former is significantly higher than the latter, which indicates that the information propagates much faster through V2V communication, as compared to vehicular mobility alone.
Figure~\ref{fig:roadblocks_result} also shows that the difference in the extent to which information has been propagated becomes greater as the communication rate $\beta$ increases, since the larger the $\beta$ value, the greater the impact of V2V on information propagation.

\subsection{Initial locations of informed vehicles} \label{initial_location}
\begin{figure}[htp]
	\centering
	\hspace{20pt}
	\subfloat[CBD area]{%
		\includegraphics[width=.3\linewidth]{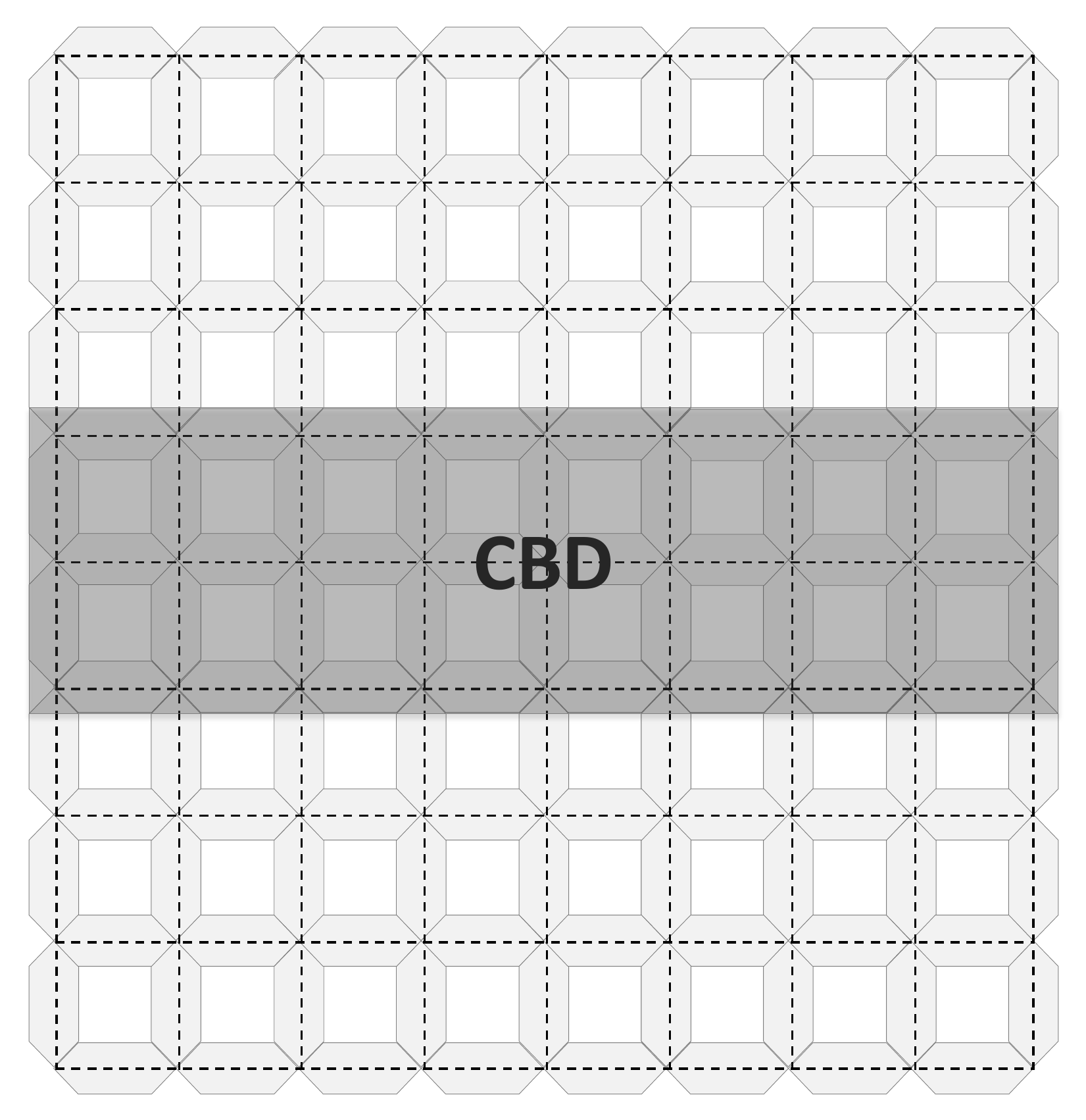}
		\label{fig:loc_grid_a}
	}%
	\hspace{30pt}
	\subfloat[Fraction of informed vehicles]{%
		\includegraphics[width=.4\linewidth]{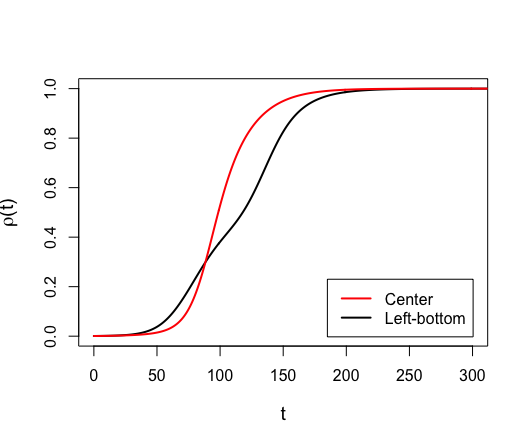}
		\label{fig:loc_grid_gamma0.5}
	}%
	\captionsetup{singlelinecheck = false, justification=raggedright, labelsep=space}
	\caption{(a) Clustered grid road topology. The CBD area is represented by a dark shade. (b) The fraction of informed vehicles over time for $\gamma=0.5$ (evening rush hour). The red lines represent the fraction of informed vehicles when the information propagates from the center, and the black lines represent the fraction of informed vehicles when the information propagates from the bottom-left corner.}
\end{figure}
\begin{figure}[htp]
	\centering
	\subfloat{%
		\includegraphics[width=.18\linewidth]{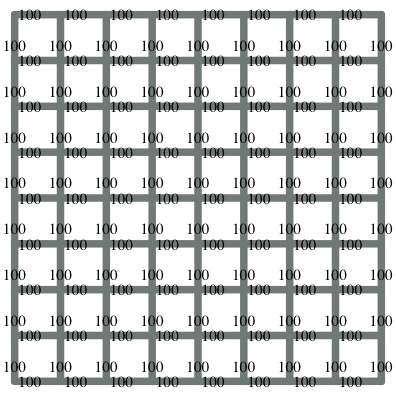}
		\label{ini_loc_geo_gamma0.5_a}
	}%
	\subfloat{%
		\includegraphics[width=.18\linewidth]{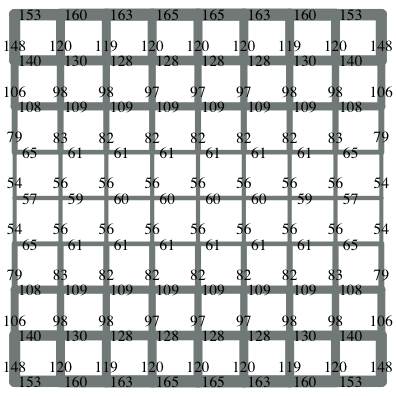}
		\label{ini_loc_geo_gamma0.5_b}
	}%
	\subfloat{%
		\includegraphics[width=.18\linewidth]{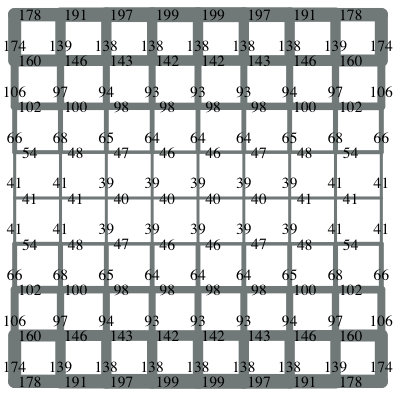}
		\label{ini_loc_geo_gamma0.5_c}
	}%
	\subfloat{%
		\includegraphics[width=.18\linewidth]{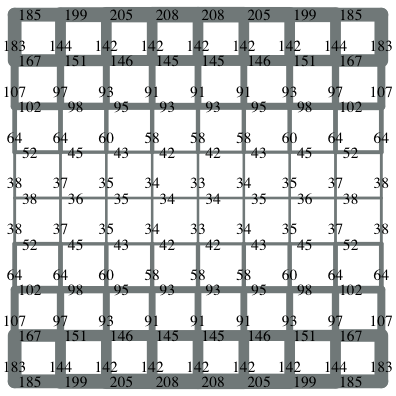}
		\label{ini_loc_geo_gamma0.5_d}
	}%
	\subfloat{%
		\includegraphics[width=.18\linewidth]{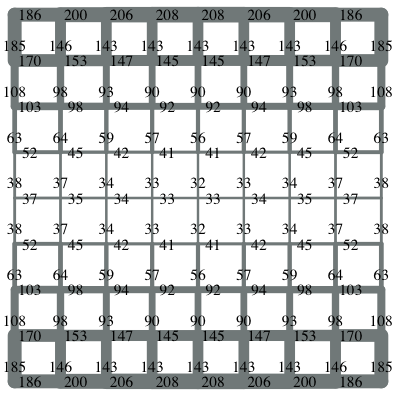}
		\label{ini_loc_geo_gamma0.5_e}
	}%
	\\
	\vspace{-7pt}
	\subfloat{%
		\includegraphics[width=.18\linewidth]{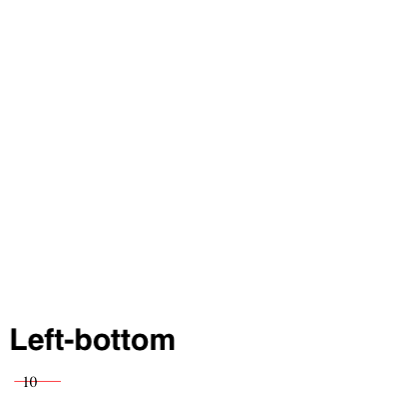}
	}%
	\subfloat{%
		\includegraphics[width=.18\linewidth]{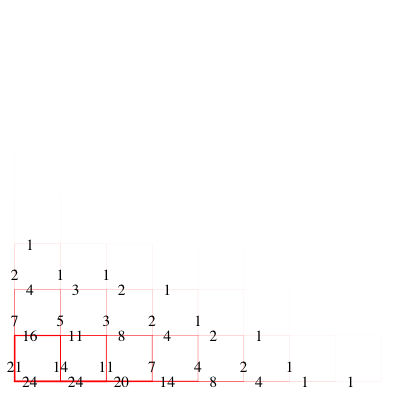}
	}%
	\subfloat{%
		\includegraphics[width=.18\linewidth]{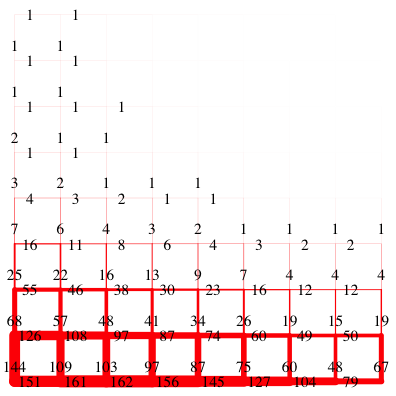}
	}%
	\subfloat{%
		\includegraphics[width=.18\linewidth]{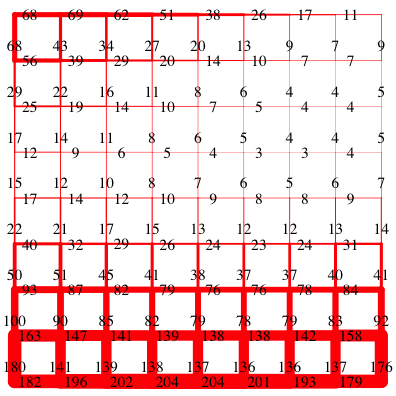}
	}%
	\subfloat{%
		\includegraphics[width=.18\linewidth]{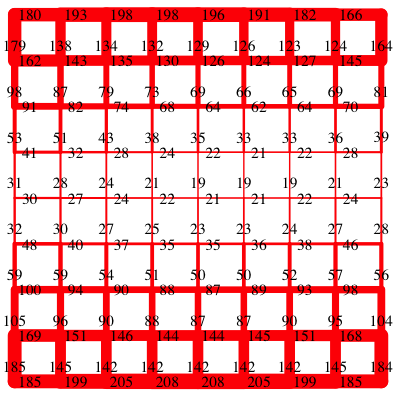}
	}%
	\\
	\vspace{-7pt}
	\addtocounter{subfigure}{-10}
	\subfloat[t=0s]{%
		\includegraphics[width=.18\linewidth]{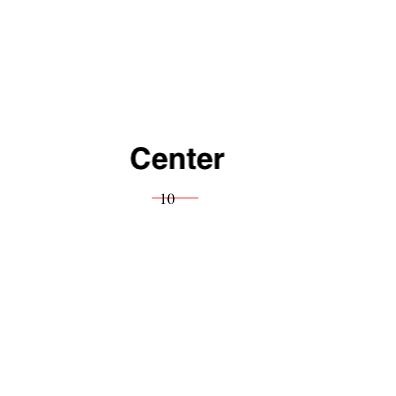}
	}%
	\subfloat[t=40s]{%
		\includegraphics[width=.18\linewidth]{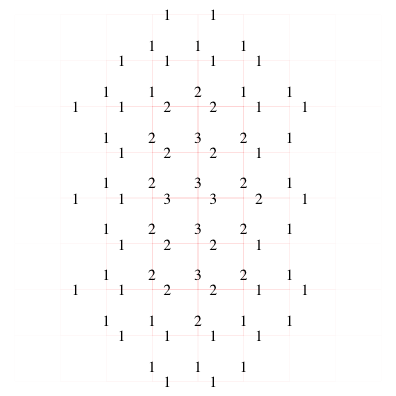}
	}%
	\subfloat[t=80s]{%
		\includegraphics[width=.18\linewidth]{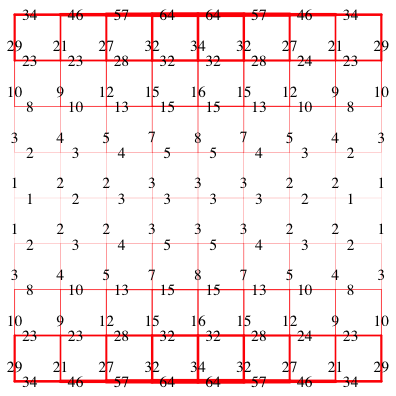}
	}%
	\subfloat[t=120s]{%
		\includegraphics[width=.18\linewidth]{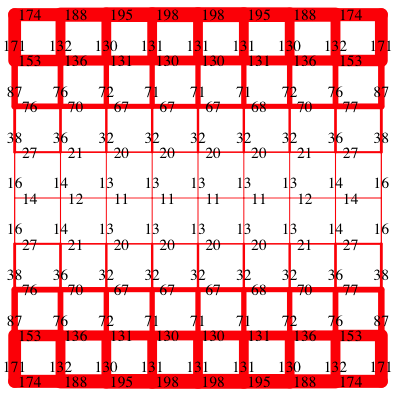}
	}%
	\subfloat[t=160s]{%
		\includegraphics[width=.18\linewidth]{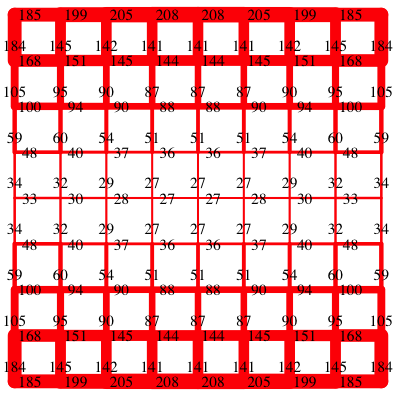}
	}%
	\captionsetup{singlelinecheck = false, justification=raggedright, labelsep=space}
	\caption{The first row of gray represents the number of vehicles in each road segment over time, and the second and third rows of red represent the number of vehicles informed in each road segment over time when information is propagated from 10 vehicles located in the lower left cluster and in the center cluster respectively. The thickness of the road segment is linearly proportional to the number of vehicles, that is, when the thickness corresponding to one vehicle is $x$, the thickness of the vehicle $v$ is expressed by $v\cdot x$. The number indicates the number of vehicles located in each road segment.} 
	\label{ini_loc_geo_gamma0.5}
\end{figure}

In the previous two examples, we have respectively considered that the unexpected event and the road obstructions occur in the center of the city. However, these can occur in anywhere in the city, so we investigate the impact of the location of initially informed vehicles on information propagation in this subsection. In particular, the impact of the location of initially informed vehicles on information propagation is expected to be more accentuated in habitations that are limited by natural boundaries such as coastal areas (e.g., Hong Kong and Manhattan). Thus, one might think that the message propagation is accelerated if the initially informed vehicles are centrally located because information can propagate simultaneously in all directions towards the boundary; using our framework, we will show that this is not necessarily the case even under the condition that vehicles are uniformly distributed in the entire region at an initial time (equivalently, no initial disparity of traffic density). This is because the information propagation speed is also affected by the vehicle movement pattern; we show this counter-intuitive phenomenon through the following example. As an example of a coastal area, we consider the finite grid area like Manhattan and assume that the CBD is located in the middle of the city (Figure~\ref{fig:loc_grid_a}). We set the parameters to $\beta = 10$, $\lambda = 0.05$. The total number of vehicles $N=14400$ is uniformly distributed at the initial time. We consider the evening rush hour mobility pattern with $\gamma = 0.5$. When information is propagated from the lower left cluster located in the periphery, information propagates rapidly to vehicles located at the bottom of the periphery where traffic density is increasing as can be seen in Figure~\ref{ini_loc_geo_gamma0.5_b}. Hence, at an early stage, information propagates faster when it propagates from the lower left corner (Figure~\ref{fig:loc_grid_gamma0.5}). However, from a certain point onwards, information propagates more quickly when information propagates from the center. When the information propagates from the bottom left corner, it takes time until the information is delivered to the top area of the periphery where traffic density is increasing as can be seen in the second row of Figure~\ref{ini_loc_geo_gamma0.5_c}. This counter-intuitive phenomenon shows that the effect of the initial location of informed vehicles on the dynamics of the information propagation can vary significantly depending on the temporal variation of traffic density.

\begin{remark}
The mathematical framework we have considered in Section~\ref{sec:2} applies to the propagation of any information in V2V network: useful information (i.e., traffic congestion), and malicious content (i.e., malware). The numerical validation in Section~\ref{sec:3} also applies both kinds of messages. However, the examples in this section (Section~\ref{sec:4}) pertain to desirable information. Specifically, Section~\ref{unexpected_events} and \ref{roadblocks} are essentially investigating if information travels faster in V2V network using both communication and mobility as opposed to using only mobility. In Section~\ref{initial_location}, we investigate the impact of the location of initially informed vehicles on the propagation of information. Both of these would also have a bearing on the spread of undesirable messages such as malware. Not all the messages that travel on a shared vehicle network will be benign; with our models, shared transportation companies can also consider their vulnerability to cyberattacks and begin to devise plans to protect against intrusions. Many of these attributes constitute directions for future research. 
\end{remark}

\section{Computation time} \label{sec:5}

A significant benefit of our model is that it is computationally tractable regardless of the number of vehicles. The statistics of the information propagation converge to the solution of the differential equations as the number of vehicles goes to infinity. Thus, the result becomes more accurate as the number of vehicles increase, and the computation time to numerically solve the differential equations does not depend on the number of vehicles. The number of variables and the number of differential equations are linear (twice) in the number of clusters. Nonetheless, we show that the computation time is still tractable even for a large number of clusters. 

We report the computation time using a computer which is not computationally high-end; a 2.8 GHz Intel Core i7 processor and 16 GB of RAM. We solved the equations using the \texttt{ode} function from the \textbf{deSolve} package of R with lsoda integration method developed by \cite{petzold1983automatic}. We solve the equations for 200 seconds, producing output every 1 second. We consider the size of a cluster to be the V2V communication range so as to enable all vehicles in the same cluster to communicate with each other. Note that the V2V communication messages have a range over 300 meters \citep{v2v_comm}. We consider extensive, complex transportation networks with relatively long roads and calculate the number of clusters, and then obtain the computation time. First, a road of approximately 366 km from New York City to Washington, D.C. via the Highway I-95S can consist of approximately 1220 clusters. The computation time required to numerically solve the differential equations for the 1220 clusters is only about 150 seconds. Next, as a more challenging example, consider the entire metropolitan region. The total road network size in Hong Kong is 2107 km \citep{hk_roadsize}, which can be covered with 7024 clusters. The computation time under a one-dimensional road consisting of 7024 clusters is approximately 172 minutes. Lastly, US Highway 101 is a North-South highway that runs through California, Oregon, and Washington on the West Coast, with a total length of nearly 2,500 km. This entire West coast highway can be covered with 8334 clusters, and the computation time is 240 minutes. These examples show that information propagation between vehicles can be identified within a reasonable time frame even in extreme cases using clustered epidemiological differential equation model.

\section{Conclusion} \label{sec:6}
V2V technologies bridge two interconnected and interdependent infrastructures: the communications infrastructure and the transportation infrastructure (including the vehicular infrastructure, the sensor infrastructure, and the physical roadway capacity). In this manuscript, we developed a continuous-time Markov chain to describe the information propagation process through enabled vehicles. Our models converge to a solution of a set of clustered epidemiological differential equations which lend itself to fast computation. We then demonstrate the applicability of this model in various scenarios: both real world scenarios and hypothesized scenarios of outages and system perturbations. We find that our models match actual trajectory data with very little error, demonstrating the applicability of our models.

Our findings are of critical importance for shared mobility services, as they invest in and consider deployment of V2V-enabled vehicle technologies. The models presented herein inform shared mobility companies of the benefits of deploying a V2V-enabled vehicles over different transportation network typologies as a function of the temporal variation of the density of V2V-enabled vehicles and the communication condition of V2V-enabled vehicles. Companies considering shared connected vehicles can scrutinize where and how their vehicles should be deployed to share information for maximum safety and efficiency benefit. Overall, our work captures the dynamics of information propagation over a connected vehicles, enabling shared mobility services, individuals, and transportation system operators to see the benefits and drawbacks of large-scale V2V-enabled vehicle deployments.

%

\appendix
\section{Proof of the result in Section~\ref{ssec:2.A}}

The stochastic model for the information propagation can be approximated by a solution of ordinary differential equations \eqref{eqn:diff_eqn}. To this end, we first present the results developed by~\cite{kurtz1970solutions}.
\begin{definition}\label{def:density-dependent}
	One parameter family of Markov chain $X(t)$ with state spaces contained in $\mathbb{Z}^{K}$ is called density dependent if there exist continuous functions $f({\bf{x}},{\bf{h}}) : \mathbb{R}^K \times \mathbb{Z}^K \rightarrow \mathbb{R}$, such that the transition rates $q({\bf{k}},{\bf{k}}+{\bf{h}})$ from ${\bf{k}}$ to ${\bf{k}}+{\bf{h}}$ is given by 
	\begin{equation}\label{eqn:density-dependent} 
	q({\bf{k}},{\bf{k}}+{\bf{h}}) = Nf\left( \frac{{\bf{k}}}{N},{\bf{h}} \right),\quad {\bf{h}} \neq {\bf 0}.
	\end{equation}
\end{definition}
We can obtain a new Markov chain by scaling with N, and the scaled process $X_N(t)$ is defined by
\begin{equation}
X_N(t) := \frac{X(t)}{N} =\frac{1}{N}({\bf{n}}^I(t),{\bf{n}}^S(t)).
\end{equation}
If a Markov process $X(t)$ is density dependent, under certain conditions, a scaled process $X_N(t)$ can be approximated by a solution of ordinary differential equations determined by the following function 
\begin{equation} \label{eqn:F}
F({\bf x}) := \sum_{{\bf{h}}\neq {\bf 0}} {\bf h} f\left({\bf x},{\bf h} \right), 
\end{equation}
which is the limiting mean increment.
The following result \citep{kurtz1981approximation,ethier2009markov} provides sufficient conditions for convergence of the scaled process $X_N(t)$ to the unique trajectory of an deterministic path when $N$ is large.

\begin{existing result} \label{thm:kurtz}
	Suppose for each compact $E \in \mathbb{R}^{K}$ 
	\begin{equation} \label{eqn:thm_cond1}
 		\sum_{{\bf h}} 	|{\bf h}| \sup_{{\bf x} \in E}  f\left({\bf x},{\bf h} \right)< \infty,
	\end{equation}
	and there exist $M_E>0$ such that
		\begin{equation} \label{eqn:Lipschitz}
	|F({\bf x})-F({\bf y})|<M_E|{\bf x}-{\bf y}|\quad {\bf x},{\bf y}\in E.
	\end{equation}
	Suppose $\lim_{N \to \infty} X_N(0)= {\bf x}(0)$, and ${\bf x}(t)$ satisfies 
	\begin{equation} \label{eqn:ode}
	{\bf x}(t)={\bf x}(0)+\int_{0}^{t} F({\bf x}(s)) ds,
	\end{equation}
	for all $t\geq 0$ (in particular $\sup_{s\leq t}|{\bf x}(s)|<\infty$). Then
	\begin{equation}
	\begin{split}
	\lim_{N \to \infty}\sup_{s\leq t} \left|X_N(s)-{\bf x}(s)\right|=0\quad \text{a.s.  for all } t>0
	\end{split}       
	\end{equation} 
\end{existing result}
We now use this result to approximate the dynamics of infomation propagation through the solution of ordinary differential equations. Recall a set ${E}:=\left\{({\bf{I}},{\bf{S}})\; |\; I_i\geq 0, \; S_i\geq 0 : i=1,2,...,J,\; \sum_{i=1}^{J} (I_i+S_i) =1 \right\}$ with $({\bf{I}},{\bf{S}})=(I_1,I_2,...,I_J,S_1,S_2,...,S_J)$. Note that $S^N/N$ is a subset of ${E}$ and the scaled process $X_N(t)$ is contained in ${E}$. Also note that ${\bf{I}}$ and ${\bf{S}}$ respectively have physical connotations of fraction of informed and non-informed vehicles in each cluster as discussed. Let the function $ f\left( {\bf x},{\bf{h}} \right)$, ${\bf x}\in E$, ${\bf{h}}\in \mathbb{Z}^{2J}$, be defined as 
\begin{equation} \label{eqn:f_function}
f\left({\bf x},{\bf{h}}\right) =
\begin{cases}
\lambda^I_{jk}\left({\bf{x}}\right) \cdot I_j, \quad & {\bf{h}}=- {\bf 1}_j + {\bf 1}_k,\; j\neq k\\
\lambda^S_{jk}\left({\bf{x}}\right) \cdot S_j, \quad & {\bf{h}}=- {\bf 1}_{J+j} + {\bf 1}_{J+k}, \; j\neq k\\
\beta_{jk} \cdot I_j \cdot S_k, \quad & {\bf{h}}= {\bf 1}_{k} - {\bf 1}_{J+k}\\
0, \quad & \text{otherwise}.
\end{cases}       
\end{equation}
where ${\bf x}=({\bf I},{\bf S})$. Since transition rate \eqref{eqn:transition_rate} can be written as the form of $q\left( {\bf k}, {\bf k} + {\bf{h}} \right) = N f\left(\frac{\bf k}{N},{\bf{h}} \right),\; {\bf{h}}\neq {\bf 0}$, the Markov process satisfies density-dependent property (Definition~\ref{def:density-dependent}). From \eqref{eqn:F} and \eqref{eqn:f_function}, the function $F({\bf x})=F({\bf I},{\bf S})$, can be expressed as 

\begin{equation} \label{eqn:2.5}
F({\bf I},{\bf S}) = \sum_{{\bf{h}}\neq {\bf 0}} {\bf{h}} f\left(({\bf{I}},{\bf{S}}),{\bf{h}} \right) =
\begin{cases}
-\sum_{k\neq 1}^{J} \lambda^I_{1k}\left({\bf{I,S}}\right) \cdot I_1 &+  \sum_{k=1}^{J}\beta_{k1}  I_k S_1    + \sum_{k\neq 1}^{J} \lambda^I_{k1}\left({\bf{I,S}}\right)\cdot  I_k \\
&\vdots\\
-\sum_{k\neq J}^{J} \lambda^I_{Jk}\left({\bf{I,S}}\right) \cdot I_J &+  \sum_{k=1}^{J}\beta_{kJ}  I_k S_J    + \sum_{k\neq 1}^{J} \lambda^I_{kJ}\left({\bf{I,S}}\right)\cdot  I_k\vspace{5pt}\\

-\sum_{k\neq 1}^{J}\lambda^S_{1k}\left({\bf{I,S}}\right) \cdot S_1 &-   \sum_{k=1}^{J}\beta_{k1}  I_k S_1    + \sum_{k\neq 1}^{J}\lambda^S_{k1}\left({\bf{I,S}}\right) \cdot S_k\\
&\vdots\\
-\sum_{k\neq J}^{J}\lambda^S_{Jk}\left({\bf{I,S}}\right) \cdot S_J &-   \sum_{k=1}^{J}\beta_{kJ}  I_k S_J    + \sum_{k\neq J}^{J}\lambda^S_{kJ}\left({\bf{I,S}}\right) \cdot S_k.
\end{cases}       
\end{equation}

\begin{lemma} \label{lemma1}
	Suppose for $i,j = 1,2,...,J$ and $i\neq j$, mobility rate functions $\lambda^I_{ij} : E \rightarrow \mathbb{R}$ and $\lambda^S_{ij} : E \rightarrow \mathbb{R}$ are bounded and Lipschitz continuous on $E$. Then, the function $F$ is Lipschitz continuous on $E$. 
\end{lemma}
\proof
	Let ${\bf{x}}=({\bf{I},\bf{S}})=(I_1,I_2,...,I_J,S_1,S_2,...,S_J))$ and ${\bf{y}}=(\bar{\bf{I}},\bar{\bf{S}})=(\bar{I}_1,\bar{I}_2,...,\bar{I}_J,\bar{S}_1,\bar{S}_2,...,\bar{S}_J))$ be points in $E$. Starting from $|F_i({\bf{x}})-F_i({\bf{y}})|$ for $i=1,2,...,J$, we have 
\begin{equation}
\begin{split}
\left|F_i({\bf{x}})-F_i({\bf{y}})\right|=&\left|-\sum_{k\neq i}^{J}\left(\lambda_{ik}^I({\bf{x}}) I_i-\lambda_{ik}^I({\bf{y}}) \bar{I}_i\right)  + \sum_{k=1}^{J}\beta_{ki} (I_k S_i-\bar{I}_k\bar{S}_i) + \sum_{k\neq i}^{J} \left(\lambda_{ki}^I({\bf{x}}) I_k-\lambda_{ki}^I({\bf{y}}) \bar{I}_k\right)\right|\\
=&\left|-\sum_{k\neq i}^{J}\lambda_{ik}^I({\bf{x}}) (I_i-\bar{I}_i)-\sum_{k\neq i}^{J} \left(\lambda_{ik}^I({\bf{x}})-\lambda_{ik}^I({\bf{y}})\right) \bar{I}_i  + \sum_{k=1}^{J}\beta_{ki} I_k (S_i-\bar{S}_i) + \sum_{k=1}^{J}\beta_{ki} \bar{S}_i(I_k-\bar{I}_k)\right.\\
&\;\left.+ \sum_{k\neq i}^{J}\lambda_{ki}^I({\bf{x}}) (I_k-\bar{I}_k)+\sum_{k\neq i}^{J} \left(\lambda_{ki}^I({\bf{x}})-\lambda_{ki}^I({\bf{y}})\right) \bar{I}_k\right|.
\end{split}
\end{equation}
for $i=1,2,...,J$. Suppose for $i,j = 1,2,...,J$ and $i\neq j$, the mobility rate function $\lambda^I_{ij}$ and $\lambda^S_{ij}$ are bounded above by $\widehat{\lambda^I_{ij}}$ and $\widehat{\lambda^S_{ij}}$ respectively. Suppose further that $\lambda^I_{ij}$  and $\lambda^S_{ij}$ are Lipschitz continuous in the sense that $|\lambda^I_{ij}({\bf{x}})-\lambda^I_{ij}({{\bf{y}}})|\leq L^I_{ij}\cdot|{\bf{x}}-{\bf{y}}|$ with Lipschitz constant $L^I_{ij}$ and $|\lambda^S_{ij}({\bf{x}})-\lambda^S_{ij}({{\bf{y}}})|\leq L^S_{ij}\cdot|{\bf{x}}-{\bf{y}}|$ with Lipschitz constant $L^S_{ij}$.
Then, we have
\begin{equation}
\begin{split}
\left|F_i({\bf{x}})-F_i({\bf{y}})\right|\;\;\leq\;\;&\sum_{k\neq i}^{J}\widehat{\lambda_{ik}^I} \left|I_i-\bar{I}_i\right|+\sum_{k\neq i}^{J} L^I_{ik}|{\bf{x}}-{\bf{y}}|  + \sum_{k=1}^{J}\beta_{ki}|S_i-\bar{S}_i| + \sum_{k=1}^{J}\beta_{ki} |I_k-\bar{I}_k|\\
&\;+ \sum_{k\neq i}^{J}\widehat{\lambda_{ki}^I} |I_k-\bar{I}_k|+\sum_{k\neq i}^{J} L^I_{ki}|{\bf{x}}-{\bf{y}}|.
\end{split}
\end{equation}
Since $\left|I_i-\bar{I}_i\right| \leq \left|({\bf{I},\bf{S}})-({\bar{\bf{I}},\bar{\bf{S}}})\right|=|{\bf{x}}-{\bf{y}}|$ and $\left|S_i-\bar{S}_i\right| \leq \left|({\bf{I},\bf{S}})-({\bar{\bf{I}},\bar{\bf{S}}})\right|=|{\bf{x}}-{\bf{y}}|$ for $i=1,2,...,J$, we have
\begin{equation}
\begin{split}
\left|F_i({\bf{x}})-F_i({\bf{y}})\right|\leq K_i\cdot|{\bf{x}}-{\bf{y}}|\quad\quad i=1,2,...,J
\end{split}
\end{equation}
where $K_i=\sum_{k\neq i}^{J}\widehat{\lambda_{ik}^I} +\sum_{k\neq i}^{J} L^I_{ik}  + 2\sum_{k=1}^{J}\beta_{ki} + \sum_{k\neq i}^{J}\widehat{\lambda_{ki}^I} +\sum_{k\neq i}^{J} L^I_{ki}$.
Through the same way, for $i=J+1, J+2,...,2J$, we have
\begin{equation}
\begin{split}
\left|F_i({\bf{x}})-F_i({\bf{y}})\right|\leq K_i\cdot|{\bf{x}}-{\bf{y}}|\quad\quad i=J+1, J+2,...,2J
\end{split}
\end{equation}
where $K_i=\sum_{k\neq i}^{J}\widehat{\lambda_{ik}^S} +\sum_{k\neq i}^{J} L^S_{ik}  + 2\sum_{k=1}^{J}\beta_{ki} + \sum_{k\neq i}^{J}\widehat{\lambda_{ki}^S} +\sum_{k\neq i}^{J} L^S_{ki}$. Thus, the component functions $F_i$ for $i=1,2,...,2J$ are Lipschitz continuous on $E$; consequently, the function $F$ is Lipschitz continuous on $E$.
\endproof
By Lemma \ref{lemma1}, the mobility rate functions $\lambda^I_{ij}$ and $\lambda^S_{ij}$ for $i,j = 1,2,...,J$ and $i\neq j$, are bounded and Lipschitz continuous on $E$, so the condition \eqref{eqn:Lipschitz} is satisfied. The state space $S^N$ is finite and the function $f({\bf x},{\bf h})$ for each ${\bf h}$ is bounded, so the condition \eqref{eqn:thm_cond1} is satisfied. Consequently, if $\lim_{N \to \infty} \frac{1}{N}({\bf{n}}^I(0),{\bf{n}}^S(0)) = \left({\bf{I}}(0),{\bf{S}}(0)\right)$, the scaled process $X_N(t)$ converges to the solution of the ordinary differential equations \eqref{eqn:diff_eqn} as the total number of vehicles $N$ increases.



\bibliographystyle{elsarticle-harv} 
\bibliography{MyLib}


\end{document}